\documentclass[a4paper,13pt]{article}
\usepackage{amsmath,epsfig}
\usepackage{subfigure}
\usepackage{amssymb,amsfonts}
\usepackage{latexsym}
\usepackage{epsfig}
\newbox\pippobox

\usepackage{amsfonts,amsmath,amsthm,amssymb}
\usepackage{txfonts}
\usepackage{mathrsfs}
\usepackage{color}
\usepackage{slashed}
\usepackage{epsf}

\newcommand{\el}[1]{\label{#1}}
\newcommand{\er}[1]{\eqref{#1}}
\newcommand{\ci}[1]{}
\newcommand{\ke}{\rangle}
\newcommand{\br}{\langle}
\newcommand{\lb}{\left(}
\newcommand{\rb}{\right)}
\newcommand{\lc}{\left.}
\newcommand{\rc}{\right.}
\newcommand{\lsb}{\left[}
\newcommand{\rsb}{\right]}

\newcommand{\nn}{\nonumber \\}

\newcommand{\ba}{\begin{eqnarray}}
\newcommand{\ea}{\end{eqnarray}}
\newcommand{\bea}{\begin{eqnarray}}
\newcommand{\eea}{\end{eqnarray}}
\newcommand{\be}{\begin{equation}}
\newcommand{\ee}{\end{equation}}
\newcommand{\bal}{\begin{align}}
\newcommand{\eal}{\end{align}}
\newcommand{\bay}[1]{\left(\begin{array}{#1}}
\newcommand{\eay}{\end{array}\right)}

\newcommand{\ie}{\textit{i.e.}, }
\newcommand{\iv}[1]{{#1}^{-1}}
\newcommand{\st}[1]{|#1\ke}

\newcommand{\zt}[1]{\textrm{#1}}

\newcommand{\no}{\nonumber \\}
\newcommand{\lla}{\langle}
\newcommand{\llb}{\rangle}
\newcommand{\ep}{\epsilon}

\newcommand{\f}{\frac}
\def\xa{{\alpha}}

\def\ddd{\cdot\cdot\cdot}

\def\xD{{\Delta}}
\def\xe{{\epsilon}}

\def\xG{{\Gamma}}

\def\xS{{\Sigma}}

\def \Tr {{\rm Tr}}

\def\CC{{\cal C}}

\def\BR{\mathbb{R}}
\def\BS{\mathbb{S}}

\title{Conformal bootstrap to R\'enyi entropy in 2D Liouville and super-Liouville CFTs}          % Enter your title between curly braces
\author{Song He$^{1,2}$\footnote{hesong17@gmail.com}}
\date{}
\begin{document}

\maketitle

\begin{center}
{\it
$^{1}$ Max Planck Institute for Gravitational Physics (Albert Einstein Institute)
Am M\"{u}hlenberg 1, 14476 Golm, Germany\\
\vspace{2mm}
$^{2}$Center for Theoretical Physics and College of Physics, Jilin University, Changchun 130012, People's Republic of China\\
}
\vspace{10mm}
\end{center}

\begin{abstract}
The R\'enyi entanglement entropy (REE) of the states excited by local operators in two-dimensional irrational conformal field theories (CFTs), especially in Liouville field theory (LFT) and $\mathcal{N}=1$ super-Liouville field theory (SLFT), has been investigated. In particular, the excited states obtained by acting on the vacuum with primary operators were considered. {We start from evaluating the second REE in a compact $c=1$ free boson field theory at generic radius, which is an irrational CFT. Then we focus on the two special irrational CFTs, e.g., LFT and SLFT. In these theories, the second REE of such local excited states becomes divergent in early and late time limits. For simplicity, we study the memory effect of REE for the two classes of the local excited states in LFT and SLFT. In order to restore the quasiparticles picture, we define the difference of REE between target and reference states, which belong to the same class. The variation of the difference of REE between early and late time limits always coincides with the log of the ratio of the fusion matrix elements between target and reference states. Furthermore, the locally excited states by acting generic descendent operators on the vacuum have been also investigated. The variation of the difference of REE is the summation of the log of the ratio of the fusion matrix elements between the target and reference states, and an additional normalization factor. Since the identity operator (or vacuum state) does not live in the Hilbert space of LFT and SLFT and no discrete terms contribute to REE in the intermediate channel, the variation of the difference of REE between target and reference states is no longer the log of the quantum dimension which is shown in the 1+1-dimensional rational CFTs (RCFTs).}
\end{abstract}

%\keywords{R\'enyi Entanglement Entropy, Liouville field theory, Conformal boostrap.}

%\begin{document}

%\maketitle
% Set the beginning of a LaTeX documen

%\newpage
%\tableofcontents
\baselineskip 18pt
\thispagestyle{empty}

\newpage

\section{Introduction}
One can define some observables to detect the property of the vacuum or excited states in a local quantum field theory. For example, entanglement entropy (EE) and the R\'enyi entanglement entropy (REE) are helpful quantities to study global or non-local structures in QFTs. For a sub-system $A$, both of them are defined as a function of the reduced density matrix $\rho_A$ which can be obtained by tracing out the degrees of freedom of the complementary of $A$ in the original density matrix $\rho$.

One might wonder whether there is a kind of topological contribution to the entanglement entropy even for gapless theories, e.g., conformal field theories (CFTs) (for example, computing topological contributions in entanglement entropy called topological entanglement entropy \cite{wen} can quantify some topological properties.) Authors of the Earlier work \cite{FFN} pointed out a connection between the topological entanglement entropy and boundary entropy. Furthermore, the connection between the boundary entropy and entanglement entropy was explored in \cite{Calabrese:2004eu}. {Previously, the authors of \cite{He:2014mwa} found that the entanglement entropy of local excited states has a connection with the quantum dimension in rational CFTs. In this paper, we would like to check whether the R\'enyi and von Neumann entropies of locally excited states are still topological quantities or not in two-dimensional irrational CFTs.}

{The $n$th R\'enyi entanglement entropy $S^{(n)}_A$ is defined by $S^{(n)}_A=\log\mbox{Tr}[\rho_A^n]/(1-n)$. By analytical continuation of $n$, the $S^{(n)}_A$ coincides with the von Neumann entropy in the limit $n\to 1$.} By using the so-called standard replica trick, one can calculate the entanglement entropy in field theory. One can extend \cite{Calabrese:2004eu} from vacuum states to locally excited states in CFTs. The computations of entanglement entropies for local excited states have been carried out in \cite{Alcaraz:2011tn}\cite{Palmai:2014jqa}\cite{Nozaki:2014hna} in various dimensional field theories. The entanglement entropy for free scalar fields has been investigated in \cite{Nozaki:2014hna}\cite{Nozaki:2014uaa}\cite{Shiba:2014uia}. In large $n$ CFTs with holographic dual, the entanglement entropy for locally excited states has been studied in \cite{Caputa:2014vaa}\cite{Asplund:2014coa}. This study mainly focuses on the variation of $S^{(n)}_A$ between excited states and a reference state, where the excited states are acquired by acting primary or descendent fields on the vacuum in irrational CFTs. The variation of $n$th REE is denoted by $\Delta S^{(n)}_A$.

In 2D rational CFTs, it was found \cite{He:2014mwa} that, for the locally primary excited states, the variation of $n$th R\'enyi entanglement entropy is related to the quantum dimension \cite{Moore:1988uz}\cite{Verlinde:1988sn} of the associated primary operator. The quantum dimension is the measure of the effective degrees of freedom of a local operator and it is a kind of topological quantity. In various dimensional CFTs, REE has been {studied in \cite{He:2014gva, Guo:2015uwa, Nozaki:2015mca, Caputa:2016tgt, Chen:2016kyz, Nozaki:2016mcy, Mertens:2016tqv, Zhou:2016ykv, Zhou:2016ekn, Caputa:2016yzn, Lin:2016dxa, Shiba:2017vsr, Caputa:2017tju, He:2017vyf, Jahn:2017xsg, He:2017txy, Balasubramanian:2017fan, Luo:2017ksc, Wen:2017smb, Kusuki:2017upd, Guo:2018lqq,Apolo:2018oqv}} from various perspectives. The papers \cite{Nozaki:2015mca}\cite{Caputa:2016tgt}\cite{Nozaki:2016mcy}\cite{Mertens:2016tqv} mainly concentrated on entanglement entropy in higher-dimensional field theory. The authors of \cite{Chen:2016kyz} have found the REE of local excited states in large central charge $1+1$ dimensional CFTs from holography. In particular, \cite{He:2014gva}\cite{Mertens:2016tqv} have provided a perspective from which to study R\'enyi entanglement entropy from string theory and it provides us with the one-loop correction to the large black hole entropy. In \cite{Zhou:2016ykv} and \cite{Zhou:2016ekn}, the entanglement entropy of a local excited state in some specific quantum Lifshitz models has been presented. More recently, the authors of \cite{Numasawa:2016kmo} mainly studied the local states of the product form of local operators in rational CFTs and they found that the variation of REE is consistent with the scattering process during entanglement propagation in RCFTs.

In this research, the previous study \cite{He:2014mwa}\cite{Guo:2015uwa}\cite{Chen:2015usa} on the R\'enyi entanglement entropy for the primary and descendent states has been generalized to irrational CFTs, especially for Liouville field theory (LFT) and super Liouville field theory (SLFT). {Previously, the authors of \cite{Asplund:2015eha} studied the memory effect of REE in a compact $c=1$ free boson theory at a generic radius, which is an irrational CFT.} There are two main motivations driving the research in LFT and SLFT. The first one is the representation of the spectra will be infinite dimensional in irrational CFTs, therefore, extracting entanglement entropy for local excited states will be highly nontrivial. A priori, one cannot expect that the variation of REE will still be the logarithmic of the quantum dimension. Furthermore, the quantum dimension of a local primary operator in irrational CFTs will be quite different from that in the 1+1-dimensional rational CFTs. Discovering how to measure variation of REE in irrational CFTs in a precise, robust way is our main aim. The second is that LFT can be reformulated as 3D Chern Simons theory \cite{Harlow:2011ny} or 3D gravity theory. In the large central charge limit, the Liouville field theory might have AdS/CFT-like connections \cite{Verlinde:1989ua}\cite{McGough:2013gka}\cite{Jackson:2014nla} with 3D gravity. Basically, the boundary conditions in Chern-Simons theory are associated with the Virasoro conformal blocks. The Liouville primary fields can be regarded as monodromy defects, which was proposed by \cite{Gaiotto:2011nm}. To understand whether or not these connections are AdS/CFT-like, we would like to work out the large central charge properties of local excited states by primary fields in LFT or SLFT; because EE and REE can be probed on the field theory side and the holographic side, both of them will be good objects with which to test the properties of these connections. In this sense, the large $c$ universal properties from these data can be generated to compare with the holographic expectation \cite{Caputa:2014vaa} of REE.

{In this paper, we evaluate the second REE in a compact $c=1$ free boson field theory at a generic radius, which is an irrational CFT, as a preliminary exercise to test the memory effect of REE. Then we mainly study the $1+1$-dimensional LFT and SLFT to show how to extract the variation of REE for locally excited states between the early time limit and the late time limit. The second REE of local primary excited states by using CFT techniques is shown in a precise way, then these calculations can be extended to the $n$th REE of primary and descendent states following \cite{He:2014mwa}\cite{Chen:2015usa}.} {From these studies, we find the REE of local excited states in LFT and SLFT is divergent which is consistent with classifications of local operator in LFT \cite{Seiberg:1990eb}. By choosing an appropriate reference state $V_{\alpha_r}|0\rangle$, we redefine a new quantity $\Delta S_{EE}^{(n)}\big[ V_{\alpha}|0\rangle, V_{\alpha_r}|0\rangle\big]$ as the difference of REE between target and reference states to measure the time evolution of REE in LFT and SLFT which is consistent with quasiparticle picture given in rational CFTs. This difference of REE $\Delta S_{EE}^{(n)}\big[ V_{\alpha}|0\rangle, V_{\alpha_r}|0\rangle\big]$ can be reduced to the $\Delta S_{EE}^{(n)}$ \cite{He:2014mwa} given in rational CFTs by choosing vacuum state as reference state. Finally, the variation of the difference of REE $\Delta S_{EE}^{(n)}\big[ V_{\alpha}|0\rangle, V_{\alpha_r}|0\rangle\big](t\to\infty)-\Delta S_{EE}^{(n)}\big[ V_{\alpha}|0\rangle, V_{\alpha_r}|0\rangle\big](t\to 0)$ depends on the ratio of the fusion matrix elements associated with $V_{\alpha}, V_{\alpha_r}$ in LFT and SLFT, unlike that of rational CFTs, which cannot be identified with the quantum dimension. That is to say, $\Delta S_{EE}^{(n)}\big[ V_{\alpha}|0\rangle, V_{\alpha_r}|0\rangle\big](t\to\infty)-\Delta S_{EE}^{(n)}\big[ V_{\alpha}|0\rangle, V_{\alpha_r}|0\rangle\big](t\to 0)$  will depend on the details of LFT and SLFT.}

The layout of this paper is as follows: in Section 2, we give the 1+1-dimensional setup and study the second REE in a precise way in LFT and $\mathcal{N}=1$ SLFT. The difference of REE between the target state and reference state $\Delta S_{EE}^{(2)}\big[ V_{\alpha}|0\rangle, V_{\alpha_r}|0\rangle\big]$ has been calculated. In Section 3, we extend above calculation to $n$-th REE in LFT to show that the variations of REE are the logarithmic of the fusion matrix elements ratio, which are quite different from those in the rational CFTs. In Section 4, the difference of REE $\Delta S_{EE}^{(n)}\big[ V_{\alpha}|0\rangle, V_{\alpha_r}|0\rangle\big]$ between states generated by acting descendent operators on the vacuum state in this setup is studied. Finally, we devote Section 5 to the conclusions and discussions and also mention some likely future problems. In the appendixes, we list some relevant notations and techniques which are necessary to our analysis.
\section{The Second R\'enyi Entanglement Entropy}

\subsection{Setup in 2D CFT}
An excited state is defined by acting an operator $\mathcal{O}_a$ on the vacuum $|0\rangle$ in a two dimensional CFT. The operator can be primary or descendent. We can make use of the Euclidean formulation and introduce the complex coordinate $(w,\bar{w})=(x+i\tau,x-i\tau)$ on $R^2$ such that $\tau$ and $x$ denote Euclidean time and space, respectively. We introduce operator ${O}_a$ at $x=-l<0$ initially and investigate its real time-evolution from time $0$ to $t$ under the Hamiltonian $H$. We develop the set-up shown in Fig. 1 and the corresponding density matrix reads as follows:
\ba
\rho(t)&=&{C_a}\cdot e^{-iHt}e^{-\epsilon H}{O}_a(-l)|0\rangle\langle 0|{{O}}^{\dagger}_a(-l)e^{-\epsilon H}e^{iHt} \nonumber\\
&=& {C_a}\cdot{O}_a(w_2,\bar{w_2})|0\rangle\langle 0|{{O}}^{\dagger}_a(w_1,\bar{w}_1),
\ea
where $C_a$ is determined by requiring Tr $\rho(t)=1$.
Here we can define coordinates as
\ba
&& w_1=i(\epsilon -it)-l, \ {}{}{} \ w_2 = -i(\epsilon+it)-l,   \label{wco}\text{} \\
&& \bar{w}_1=-i(\epsilon-it)-l,\ {}{}{}{}{}{} \ \bar{w}_2=i(\epsilon+it)-l.  \label{wcor}
\ea
$\epsilon$ is an infinitesimal positive parameter as an ultraviolet regulator. Until the end of the calculations, we treat
$\epsilon\pm it$ as purely imaginary numbers as in \cite{Nozaki:2014hna, Nozaki:2014uaa,He:2014mwa}.

To calculate variation of $n$th REE $\Delta S^{(n)}_A$, we employ the replica method in the path-integral formalism\footnote{{More precisely, the replica method for the local operator excited states in field theory have been explicitly shown in section 2.2 of \cite{Numasawa:2016kmo}.}} by generalizing the formulation for the ground states \cite{Calabrese:2004eu} to excited states \cite{Nozaki:2014hna}.
In this paper, we choose the subsystem $A$ to be an interval $0\leq x\leq L$ at $\tau=0$. For simplification, we only consider $L\to \infty$ throughout.
It leads to a $n$-sheeted Riemann surface $\Sigma_n$ with $2n$ operators ${\mathcal O}_a$ inserted.
\begin{figure}[h]
\begin{center}
\epsfxsize=6.5 cm \epsfysize=6.5 cm \epsfbox{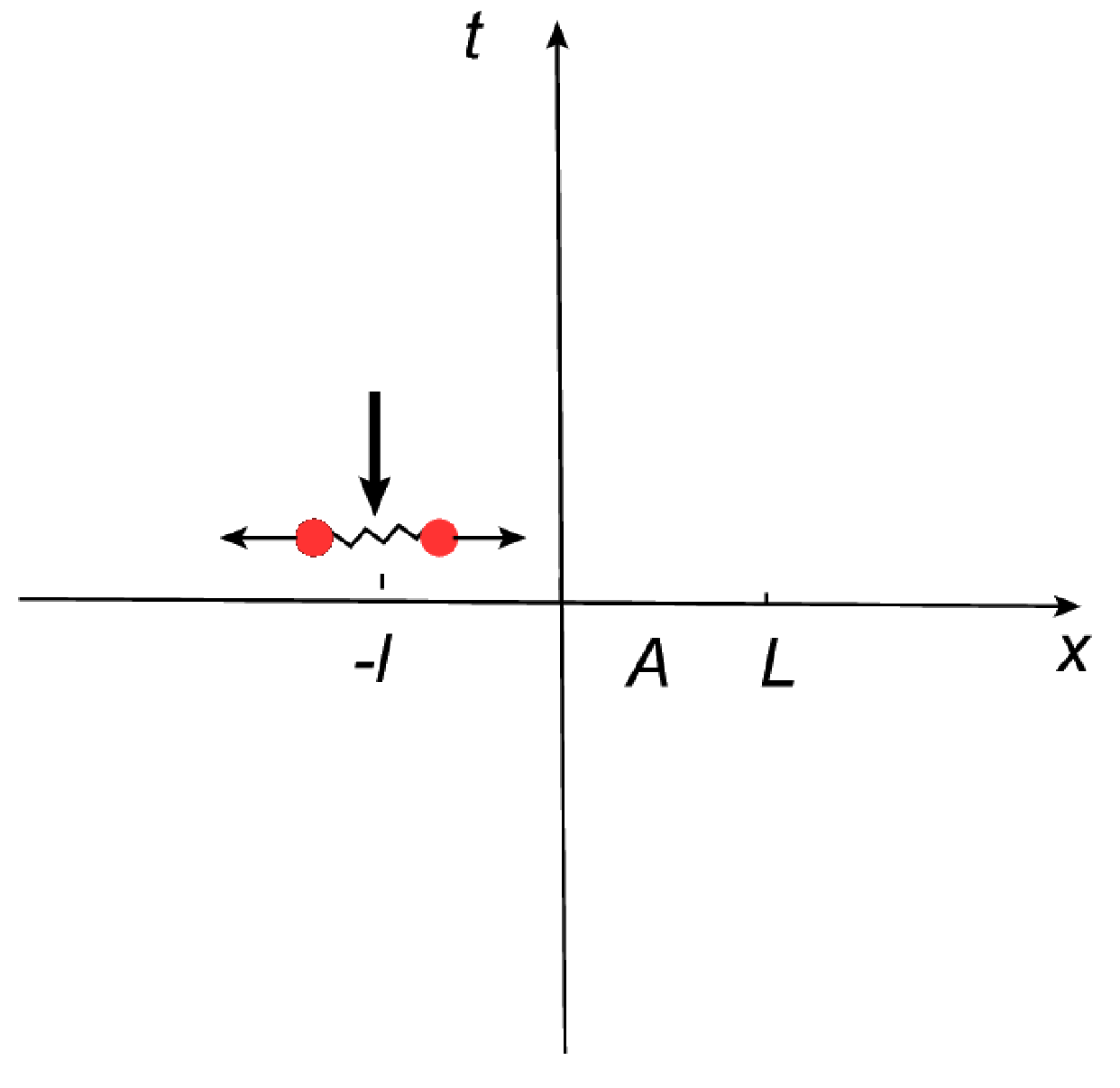}
\end{center}
\caption{This figure is to show our basic setup in two dimensional plane $ w = x + it$. We consider the subsystem A $0 <x<L$ with $L\to \infty$. We just put the local operators at $x = -l, t=0$. The local operator will trigger left- and right-moving quasi-particles with time evolution.} \label{figuresetup}
\end{figure}

Finally, the $\Delta S_A^{(n)}$ can be calculated as
\ba
\Delta S_A^{(n)}
&=&\!{1\over1-n} \Biggl[\log{\left\langle{{O}}^{\dagger}_a(w_1,\bar{w}_1){O}_a(w_2,\bar{w}_2)
 \cdots {O}_a(w_{2n},\bar{w}_{2n})\right\rangle_{\Sigma_n}}\nonumber\\
&& {}{}{}{} -n\log\left\langle{{O}}^{\dagger}_a(w_1,\bar{w}_1)
{O}_a(w_2,\bar{w}_2)\right\rangle_{\Sigma_1}\Biggr], \label{replica1}
\ea
where $(w_{2k+1},w_{2k+2})$ for $k=1,2,...,n-1$ are $n-1$ replicas of $(w_1,w_2)$ in the $k$-th sheet of $\Sigma_n$. The term in the first line in Eq.(\ref{replica1}) is given by a $2n$-point correlation function on $\Sigma_n$. Here $\Delta_a$ is the (chiral and anti-chiral) conformal dimension of the operator ${O}_a$. {One should note that $\Delta S_A^{(n)}$ can be well defined once the vacuum state belongs to the Hilbert space. That is to say one can choose the vacuum state as a good reference state to measure $\Delta S_A^{(n)}=\Delta {S}^{(n)}_{A}\big[ V_{\alpha}|0\rangle,|0\rangle\big]$ between excited states and vacuum state in rational CFTs\cite{He:2014mwa}. Otherwise, one has to choose an appropriate reference state $V_{\alpha_r}|0\rangle$ to measure $\Delta {S}^{(n)}_{A}\big[ V_{\alpha}|0\rangle, V_{\alpha_r}|0\rangle\big]$, e.g., in Liouville field theory\footnote{We will explain the details at the end of this Section.}. }

\subsection{Convention}

Firstly, we study $n=2$ i.e. the second R\'enyi entanglement entropy in detail. The calculation of
$\Delta S^{(2)}_A$ is reduced to four point functions in CFTs.

For $n=2$, one can connect the coordinate $w_i$ with $z_i$ by a conformal mapping $w_i=z_i^2$ which looks like
\ba
 w_1&=&i\epsilon+t-l\equiv re^{i\theta_1}=(z_1)^2, \nonumber\\
 w_2&=&-i\epsilon+t-l\equiv se^{i\theta_2}=(z_2)^2, \nonumber\\
 w_3&=&(i\epsilon+t-l)e^{2\pi i}\equiv re^{i(2\pi+\theta_1)}=(z_3)^2, \nonumber\\
 w_4&=&(-i\epsilon+t-l)e^{2\pi i}\equiv se^{i(2\pi+\theta_2)}=(z_4)^2.
\ea
Thus one can find
\ba
z_1&=&-z_3=\sqrt{w_1}=\sqrt{r}e^{i\theta_1/2}=i\sqrt{l-t-i\epsilon}, \nonumber\\
z_2&=&-z_4=\sqrt{w_2}=\sqrt{s}e^{i\theta_2/2}=i\sqrt{l-t+i\epsilon}. \label{zcor1}
\ea
If the readers are interested in finite size formula, please refer to \cite{He:2014mwa}.

We will follow a standard procedure of analytical continuation of Euclidean theory into its Lorentzian version. The most important and subtle point is that we should treat $\pm i\ep+t$ as a pure imaginary number in whole algebraic calculations. Finally, we take $t$ to be real only in the final expression of the variation of entropy.
Here we identify
\ba
(r\cos\theta_1,r\sin\theta_1{}{})=(-l,\epsilon-it),{}{}{} && (s\cos\theta_2,s\sin\theta_2)=(-l,-\epsilon-it),
\ea
which leads to
\ba
 r&=&\sqrt{l^2+(\epsilon-it)^2},\ \ s=\sqrt{l^2+(-\epsilon-it)^2}, \nonumber\\
rs &=&\sqrt{(l^2+\epsilon^2-t^2)^2+4\epsilon^2t^2},\ \ r^2+s^2=2(l^2+\epsilon^2-t^2), \nonumber\\
\cos(\theta_1-\theta_2)&=&2\cos^2\left(\frac{\theta_1-\theta_2}{2}\right)-1=
{l^2-\epsilon^2-t^2\over\sqrt{(l^2+\epsilon^2-t^2)^2+4\epsilon^2t^2}}. \label{rs1}
\ea

To get REE, we solely focus on the conformal cross ratio
\ba
z&=&{z_{12}z_{34}\over z_{13}z_{24}}={-(l-t)+\sqrt{(l-t)^2+\epsilon^2}\over 2\sqrt{(l-t)^2+\epsilon^2}},\nonumber\\
\bar{z}&=&{\bar{z}_{12}\bar{z}_{34}\over \bar{z}_{13}\bar{z}_{24}}
={-(l+t)+\sqrt{(l+t)^2+\epsilon^2}\over 2\sqrt{(l+t)^2+\epsilon^2}}, \label{zij1}
\ea
where $z_{ij}=z_i-z_j$.

It is useful to note the relationship
\ba
1-z={z_{14}z_{23}\over z_{13}z_{24}}.
\ea

We are interested in the two limits (i) $l>>t>>\epsilon$ (early time) and (ii) $t>>l>>\epsilon $ (late time) and from (\ref{zij1}) we can know that separately corresponds to
\ba
(i) && \ \  z\simeq \bar{z}\simeq {\epsilon^2\over 4l^2}\ (\to 0),\nonumber\\
(ii) && \ \  z\simeq 1-{\epsilon^2 \over 4t^2}\ (\to 1), \ \ \ \ \bar{z}\simeq {\epsilon^2\over 4t^2}\ (\to 0).
\ea
Note that the late time limit is quite non-trivial which originates from our analytical continuation of $t$.

\subsection{The Variation of second R\'enyi Entanglement Entropy}

The four point function on $\Sigma_2$ is mapped into that on $R^2$ by the conformal map
$w=z^2$. Thus we find
\ba
&& \lla O_a(w_1,\bar{w}_1)O_a(w_2,\bar{w}_2)O_a(w_3,\bar{w}_3)O_a(w_4,\bar{w}_4)\llb_{\Sigma_2} \no
& = &\prod_{i=1}^{4}\left|\f{dw_i}{dz_i}\right|^{-2\Delta}\lla O_a(z_1,\bar{z}_1)O_a(z_2,\bar{z}_2)O_a(z_3,\bar{z}_3)O_a(z_4,\bar{z}_4)\llb_{\Sigma_1} \no
& =&2^{-8\Delta}|z_1z_2z_3z_4|^{-2\Delta}\cdot \lla O_a(z_1,\bar{z}_1)O_a(z_2,\bar{z}_2)O_a(z_3,\bar{z}_3)O_a(z_4,\bar{z}_4)\llb_{\Sigma_1}\no
\label{tensortransform} &=&2^{-8\Delta}\cdot (rs)^{-2\Delta} \cdot \lla O_a(z_1,\bar{z}_1)O_a(z_2,\bar{z}_2)O_a(z_3,\bar{z}_3)O_a(z_4,\bar{z}_4)\llb_{\Sigma_1},
\ea
where $\Delta$ is the chiral conformal dimension of the operator $O_a$.

The two-point function looks like
\ba
\lla O_a(w_1,\bar{w}_1)O_a(w_2,\bar{w}_2)\llb_{\Sigma_1}
=\f{{C_{a}}}{|w_{12}|^{4\Delta}}=\f{{C_{a}}}{(2\ep)^{4\Delta}}\label{twopoint},
\ea
where $C_{a}$ represents normalization. Note that the four-point function is proportional to
$C_{a}^2$ and the $\Delta S^{(2)}_A$ is of course independent of $C_{a}$.
Thanks to the conformal symmetry, the four-point function on $R^2$ can be expressed as
\be
\lla O_a(z_1,\bar{z}_1)O_a(z_2,\bar{z}_2)O_a(z_3,\bar{z}_3)O_a(z_4,\bar{z}_4)\llb_{\Sigma_1}
=|z_{13}z_{24}|^{-4\Delta}\cdot G(z,\bar{z}), \label{gf}
\ee
where $(z,\bar{z})$  are given by (\ref{zij1}).

In the late time limit (ii), we finally find the ratio in (\ref{replica}) is expressed in terms of the four-point function on $R^2$:
\ba
 \mbox{Tr}\rho_A^2&=&\f{\lla O_a(w_1,\bar{w}_1)O_a(w_2,\bar{w}_2)O_a(w_3,\bar{w}_3)O_a(w_4,\bar{w}_4)\llb_{\Sigma_2}}{\left(\lla O_a(w_1,\bar{w}_1)O_a(w_2,\bar{w}_2)\llb_{\Sigma_1}\right)^2} \nonumber\\
& \simeq & \frac{1}{C_{a}^2}\cdot \left(\frac{\epsilon^2}{t}\right)^{4\Delta}\cdot \lla O_a(z_1,\bar{z}_1)O_a(z_2,\bar{z}_2)O_a(z_3,\bar{z}_3)O_a(z_4,\bar{z}_4)\llb_{\Sigma_1} \nonumber\\
&\simeq & \frac{1}{C_{a}^2}|z^{4\Delta}(1-z)^{4\Delta}|G(z,\bar z) \simeq \frac{1}{C_{a}^2}\cdot \left(\frac{\epsilon^2}{4t^2}\right)^{4\Delta}\cdot G(z,\bar{z}).
\label{ratio}
\ea

In rational CFTs, we can calculate $\Delta S_A^{(n)}$ between local excited states and the vacuum state as follows:
\ba
\Delta S_A^{(n)}&=&S_A^{(n)}(O_a|0\rangle)-S_A^{(n)}(1 |0\rangle)\nonumber\\&=&\f{1}{1-n}\log \left[\f{\lla O_a(w_1,\bar{w}_1)O_a(w_2,\bar{w}_2)\ddd O_a(w_{2n},\bar{w}_{2n})\llb_{\Sigma_n}}{\left(\lla O_a(w_1,\bar{w}_1)O_a(w_2,\bar{w}_2)\llb_{\Sigma_1}\right)^n}\right].\quad \label{replica}
\ea
Here $\Sigma_n$ denotes the $n-$sheeted Euclidean surface given by the metric
\ba
ds^2=d\rho^2+\rho^2(d \theta)^2,
\ea
where $\theta$ has the $2\pi n$ periodicity $\theta\sim \theta+2\pi n$.

Extra care should be taken when we generalise \er{replica} to the case of LFT and SLFT. We note that normally the vacuum expectation value of $n$ operators is defined as
\be
\label{normalized-n-point}
\br O_a(w_1,\bar{w}_1)O_a(w_2,\bar{w}_2)\ddd O_a(w_{2n},\bar{w}_{2n})\ke_{\Sigma_n} \equiv \frac {Z_n}{Z_{0n}}\,,
\ee
where $Z_n, Z_{0n}$ are the partition functions with or without operators inserted on $\xS_n$. {Following the replica method in the Euclidean path-integral formalism\footnote{The replica method for the local operator excited states in field theory have been explained explicitly in section 2.2 of \cite{Numasawa:2016kmo}.} \cite{Numasawa:2016kmo}}, we can express the reduced density as $\Tr \rho^n = Z_n/Z_1^n$. As a result $\xD S_A^{(n)}$ can be written as
\be
\xD S_A^{(n)} =\frac 1 {1-n} \lb \log \Tr \rho^n - \log \Tr \rho_0^n \rb = \frac 1 {1-n} \lb \log \frac {Z_n}{Z_{0n}}- n \log \frac {Z_1}{Z_{01}} \rb\,.
\ee
One can see that \er{replica} follows when identity operator $1$ belongs to the Hilbert space of the theory, however that in Liouville field theory, the $n$-point function is defined by the path integral and therefore it is not normalised
\be
\br V_{\xa_1}(w_1,\bar{w}_1)V_{\xa_2}(w_2,\bar{w}_2)\ddd V_{\xa_n}(w_{2n},\bar{w}_{2n})\ke_{\Sigma_n} \equiv Z_n.
\ee
Because $\alpha \in \{Q/2+ i p|  p\in  \mathbb{R}\} \bigcup \{Q>\alpha> 0\}\}$ in LFT and SFLT, one can not take all $V$'s to be the identity operator, \ie analytically continue $\xa = \frac Q 2 + ip$ to $p = i Q/2$.

In other words, \er{replica} applies to the case of Liouville field theory gives the R\'enyi entanglement entropy $S_A^{(n)}(\st{V_\xa})$
\ba\label{2REE}
&&S_A^{(n)}[V_{\xa}\st{0}](t) \nonumber\\&=& \f{1}{1-n}\log\f{\lla V^\dagger_{\bar \xa} (w_1,\bar{w}_1)V_\xa (w_2,\bar{w}_2) \dots V^\dagger_{\bar \xa} (w_{2n-1},\bar{w}_{2n-1}) V_\xa (w_{2n},\bar{w}_{2n})\llb_{\Sigma_n}}{\left(\lla V^\dagger_{\bar \xa}(w_1,\bar{w}_1)V_\xa(w_2,\bar{w}_2)\llb_{\Sigma_1}\right)^n}.
\ea
In LFT and SLFT considered in this paper, the identity operator does not belong to the Hilbert {space and the vacuum state \cite{vacuumstate}} can not be considered a good reference state\footnote{{The existence of the translation invariant normalisable vacuum is not self-consistent with classical equation of motion of Liouville field theory, which has been shown in \cite{vacuumstate}. The non-existence of $SL(2,C)$ invariant vacuum in the spectrum has important consequence that the identity operator does not belong to the whole Hilbert space; this means that the external Liouville momentum $\alpha$ cannot be vanishing.}}such as that in rational CFTs. Therefore, the $S_A^{(n)}(1|0\rangle)$ in the formula Eq.(\ref{replica}) cannot be applied in LFT and SLFT. {We can define the difference $\Delta {S}^{(n)}_{A}\big[ V_{\alpha}|0\rangle, V_{\alpha_r}|0\rangle\big](t)$ between of two excited states. Here $V_{\alpha}|0\rangle$ and $V_{\alpha_r}|0\rangle$ are the target state and reference state respectively. Alternatively, we calculate $S_A^{(n)}(O_a|0\rangle)$ in the early time and the late time limits and define the difference of REE between the two states }
\ba\Delta {S}^{(n)}_{A}\big[ V_{\alpha}|0\rangle, V_{\alpha_r}|0\rangle\big](t)=S_A^{(n)}[V_{\alpha}(t)]|0\rangle(t)-S_A^{(n)}[V_{\alpha_r}(t)|0\rangle](t)\ea to study time evolution. {For later convenience, we divide the primary operators in LFT and SLFT into two classes in terms of Liouville momentum, such as\footnote{{The main reason to choose reference and target states from the same class in LFT and SLFT is to calculate well defined quantity to restore the quasi-particle picture.}}
\bea\label{class}
%&&\nonumber\\
\alpha \text{ } \text{and} \text{ }\alpha_r&\in&\left
 \{
\begin{array}{cc}
&  \{\alpha|\alpha=Q/2+ i p,  p\in  \mathbb{R}\}\bigcup \{\alpha|Q/2>\text{Re}(\alpha)> {Q/4}\}\\&\bigcup \{\alpha|{Q/4>\text{Re}(\alpha)> {0}}\},  \\
   & \{\alpha|\text{Re}(\alpha)=Q/4, \text{Im}(\alpha)\neq 0\} \text{ }\text{ } \text{Marginal Case}. \\
\end{array}\right.
\eea  }

\subsection{The Second REE in $c=1$ free boson field theory}
To begin with analysis of REE in special irrational CFTs, the compact free boson theory with generic radius is a simple irrational theory to see the time evolution of the 2nd REE\footnote{The authors of \cite{Asplund:2015eha} called it as a memory effect of REE.}.
Following \cite{Asplund:2015eha}, the second the REE is as follows
 \bea\label{cone2nd}
 S _ { 2 } = \frac { \pi c L } { 2 \beta } - \log \left[ 2 ^ { - 2 c / 3 } ( z \overline { z } ) ^ { c / 12 } \right] ( 1 - z ) ( 1 - \overline { z } ) ] ^ { - c / 24 } Z ( \tau ( z ) , \overline { \tau } ( \overline { z } ) ) ] + \frac { c } { 2 } \log \frac { \beta } { 2 \pi \epsilon },
 \eea
%{\bea
%G _ { 2 } ( z , \overline { z } ) = \left\langle \sigma _ { 2 } ( 0 ) \tilde { \sigma } _ { 2 } ( z , \overline { z } ) \sigma _ { 2 } ( 1 ) \tilde { \sigma } _ { 2 } ( \infty ) \right\rangle
%\eea
%The two-sheeted manifold that defines these double-twist operators is in fact a torus. It follows that this correlation
%function is given by the torus partition function, and the mapping derived in [19,20] is
%\bea
%    G _ { 2 } ( z , \overline { z } ) = \left( 2 ^ { 16 } z \overline { z } ( 1 - z ) ( 1 - \overline { z } ) \right) ^ { - c / 24 } Z ( \tau , \overline { \tau } )
%\eea
where
\bea
Z ( \tau , \overline { \tau } ) = \operatorname { Tr } \exp \left[ 2 \pi i \tau \left( L _ { 0 } - \frac { c } { 24 } \right) - 2 \pi i \overline { \tau } \left( \overline { L } _ { 0 } - \frac { c } { 24 } \right) \right]. \eea
The partition function can be calculated with the torus modulus which is defined by \bea
    z = \frac { \theta _ { 2 } ( \tau ) ^ { 4 } } { \theta _ { 3 } ( \tau ) ^ { 4 } } , \quad \tau = i \frac { K ( 1 - z ) } { K ( z ) }.
\eea
The $K ( z ) = \frac { \pi } { 2 } \, _2F_1 \left( , \frac { 1 } { 2 } , 1 , z ^ { 2 } \right)  $.

For $z\to 0$, the relation is
\bea
    z = 16 \sqrt { q } + O ( q ) , \quad q \equiv e ^ { 2 \pi i \tau }.
\eea

For $z\to 1$, the relation is
\bea
  1 - z = 16 \sqrt { q ^ { \prime } } + O \left( q ^ { \prime } \right) , \quad q ^ { \prime } \equiv e ^ { - 2 \pi i / \tau }.
\eea

We can analyze the two light cone singularities of the second Reyni entropy of this theory
explicitly, via the torus partition function [30]
\bea
 Z ( R ) = \frac { 1 } { \eta ( \tau ) \overline { \eta } ( \overline { \tau } ) } \sum _ { e , m \in \mathbb { Z } } q ^ { h _ { e , m } } \overline { q } ^ { \overline { h } _ { e , m } },
\eea where $\eta$ denotes the Dedekind eta function.
 The conformal weights are as follows:
 \bea
     \begin{array} { l } { h _ { e , m } = \frac { 1 } { 2 } \left( \frac { e } { R } + \frac { m R } { 2 } \right) ^ { 2 } } \\ { \overline { h } _ { e , m } = \frac { 1 } { 2 } \left( \frac { e } { R } - \frac { m R } { 2 } \right) ^ { 2 } } \end{array}.
 \eea
Here $e,m$ are integer number. When $R^2$ is irrational, there is no degeneracy in the weights
and the theory is irrational CFTs.
In the early time limit  $\tau \rightarrow i \infty , \overline { \tau } \rightarrow -i \infty , q \rightarrow  0, \overline { q } \rightarrow 0 ^ { + }$, the partition function behaves as
\bea Z ( \tau , \overline { \tau } ) \sim q ^ { - c / 24 } \overline { q } ^ { - c / 24 } \sim 2 ^ { 2 c / 3 } z ^ { - c / 12 } \overline { z } ^ { - c / 12 }.\eea
Then the second REE is as follows:
  \bea
 S ^ {( 2 )} = \frac { \pi c L } { 2 \beta }+ \frac { c } { 2 } \log \frac { \beta } { 2 \pi \epsilon }.
 \eea
 In the late time limit $\tau \rightarrow i 0 ^ { - } , \overline { \tau } \rightarrow -i \infty , q \rightarrow  1, \overline { q } \rightarrow 0 ^ { + }$, the partition function behaves as
 \bea
     Z ( R ) = \frac { 1 } { \eta ( \tau ) \overline { \eta } ( \overline { \tau } ) } \left( 1 + q ^ { h _ { \min } } \overline { q } ^ { \overline { h } _ { \min } } + \cdots \right).
 \eea
We have used the following properties
 \bea
    \eta ( - 1 / \tau ) = \sqrt { - i \tau } \eta ( \tau ) , \quad \overline { \eta } ( \overline { \tau } ) = \sqrt { \frac { - i } { \overline { \tau } } } \overline { \eta } \left( \overline { \tau } ^ { \prime } \right).
 \eea
 Thus
 \bea
     Z ( R ) \sim q ^ { - 1 / 24 } \overline { q } ^ { - 1 / 24 }.
 \eea

Then, the late time limit of the second REE in $c=1$ compact free boson theory is
  \bea
 S^{( 2 )} = \frac { \pi c L } { 2 \beta }-\log(2^{-2c \over 3})+ \frac { c } { 2 } \log \frac { \beta } { 2 \pi \epsilon }.
 \eea
 The second term in (\ref{cone2nd}) does not vanish in the late time limit.

 Finally, the variation of second REE between the early time and the late time is
   \bea
 \Delta S ^ {( 2) } = {2 \over 3}\log 2.
 \eea Here we have chosen $c=1$.

The compact free boson theory with generic radius as `nearly rational' theory and the entanglement entropy has quasiparticle behavior, which is consistent with criterion of quasiparticle picture \cite{Asplund:2015eha}.
In the remaining parts of this paper, we will calculate REE in Liouville and super-Liouville Field theories which have different Hilbert space structures. The first difference is that the vacuum state does not be contained in the Hilbert space of LFT and SLFT and the second one is the that the spectrum in LFT and SLFT are continous. Due to these two differences, the above calculations cannot be applied to LFT and SLFT directly. We will show later what will happen to REE in LFT and SLFT in more detail.

\subsection{The Second REE in Liouville Field Theory}\label{2REELFT}
{We are mainly interested in the second REE here which is associated with 4-point function in terms of Eq.(\ref{ratio}). The 4-point functions in LFT have been reviewed in Appendix 6. {In our set-up, the interested $\langle V_{\bar\alpha} V_{\alpha} V_{\bar\alpha} V_{\alpha} \rangle_{\Sigma_1} $ is given by Eq.(\ref{4point}).}
In LFT, the four-point Green function of primary operator $V_\alpha$ in the $s$-channel can be expressed by}
\ba\label{4ptfunction}
&&\langle V_{\alpha_1} V_{\alpha_2} V_{\alpha_3} V_{\alpha_4} \rangle_{\Sigma_1}\nonumber\\ &=& |z_{13}|^{-4\Delta}|z_{24}|^{-4\Delta}G_{1234}({z,\bar{z}})\nonumber\\
&=&{1\over 2}|z_{13}|^{-4\Delta}|z_{24}|^{-4\Delta}\int_\BR {dp\over 2\pi} C(\alpha_1,\alpha_2, {Q\over 2}+i p) C(\alpha_3,\alpha_4,  {Q\over 2}-i p)\nonumber\\
&&F_{s}(\Delta_{i=1,2,3,4},\Delta_p,{z})F_{s}(\Delta_{i=1,2,3,4},\Delta_p,{\bar z}).
\ea
The $\Delta_{i}={\alpha_i}(Q-\alpha_i)$ is the conformal dimension of external Liouville momentum $\alpha_i \in \{\alpha|\alpha=Q/2+ i p, p\in  \mathbb{R}\}\bigcup \{\alpha|Q/2>\text{Re}(\alpha)> {Q/4}\}\bigcup \{\alpha|{Q/4>\text{Re}(\alpha)> {0}}\}$. The integration over intermediate momentum $p$ stands for contour integration over $p\in \mathbb{R}$. $F_{s}(\Delta_{i=1,2,3,4},\Delta_p,{z})$
and $F_{s}(\Delta_{i=1,2,3,4},\Delta_p,{\bar z})$ are the holomorphic and anti-holomorphic conformal blocks respectively. The DOZZ formulae $C(\bar\alpha,\alpha, {O\over 2}+i p)$ are given {in Appendix \ref{reviewLFT}. More} precisely, for $\alpha_i \in (0,Q)$
the $G_{1234}(z,\bar z)$ in $s$-channel can be modified and expressed as
follows
\ba
G_{1234}(z,\bar z)&=&\sum_{\alpha'_s\in D}D_{\alpha'_s}(z,\bar z)+\int{dp\over 2\pi} C(\alpha_1,\alpha_2, {\frac Q 2}+i p) C(\alpha_3,\alpha_4,  {\frac Q 2}-i p)\nonumber\\&&F_{s}(\Delta_{i=1,2,3,4},\Delta_p,{z})F_{s}(\Delta_{i=1,2,3,4},\Delta_p,{\bar z})\label{discreteterm}
\ea
Where $D$ denotes the discrete terms (\ref{discretetermLFT}) \footnote{{In our setup of REE, the four-point Green function does not involve any discrete terms.}} reviewed in the Appendix \ref{discrete} and \ref{appendix2}.

First, let us calculate the REE in the early time limit. One can make use of $s$-channel expression Eq.(\ref{4point}) for $\langle V_{\bar\alpha} V_{\alpha} V_{\bar\alpha} V_{\alpha} \rangle $
\ba
G_{\bar{\alpha}\alpha\bar{\alpha}\alpha}(z,\bar z) &=&
{1\over 2}\int_\BR'{dp\over 2\pi} C(\bar{\alpha},\alpha, {Q\over 2}+i p) C(\bar{\alpha},\alpha,  {Q\over 2}-i p)\nonumber\\&&F_{s}(\Delta_{i=\bar{\alpha},\alpha,\bar{\alpha},\alpha},\Delta_p,{z})F_{s}(\Delta_{i=\bar{\alpha},\alpha,\bar{\alpha},\alpha},\Delta_p,{\bar z})\label{earlytimecc}\\
&=&{1\over 2} \int_\BR {dp\over 2\pi}\Big[C(\bar{\alpha},\alpha, {Q\over 2}+i p) C(\bar{\alpha},\alpha,  {Q\over 2}-i p)\nonumber\\&&\bar{z}^{\Delta_{Q/2+i
p}-2\Delta_{\alpha}}(1+...)z^{\Delta_{Q/2+i
p}-2\Delta_{\alpha}}(1+...)\Big]\label{firstterm-1}
\ea We have shown the asymptotic behavior of the early time limit in the last step in (\ref{firstterm-1}).

Once we take early time limit of Eq.(\ref{firstterm-1}),
\ba
\lim_{(z,\bar z)\to (0, 0)}\langle V_{\bar\alpha} V_{\alpha} V_{\bar\alpha} V_{\alpha} \rangle_{\Sigma_1} &\simeq& |z_{13}|^{-4\Delta}|z_{24}|^{-4\Delta}{d^2f_{\alpha}(p)\over d p^2}\Big|_{p\to 0} \int_{\mathbb{R}} |z|^{-4\Delta_{\alpha}+2\Delta_{Q/2+i p} } p^2 \, d p \nonumber\\& \simeq& {\sqrt{\pi}\over 8\times 2! }{d^2f_{\alpha}(p)\over d p^2}\Big|_{p\to 0} |z|^{-2(2\Delta_{\alpha} - \Delta_{Q/2})} \ln^{-\frac32}|1/z|
\ea Where we define
\be\el{deffP} f_{\alpha}(p)=C(\bar\alpha,\alpha,{Q\over 2}+i p)C(\bar\alpha,\alpha,{Q\over 2}-i p)\,.\ee
Here we have used saddle point approximation presented in Appendix \ref{appendix2} to obtain the leading behaviour in the early time limit.
The two-point Green function for primary operator in LFT is as follows:
\ba
\langle V_{\alpha}(x_1)V_{\alpha}(x_2) \rangle_{\Sigma_1}= { S(\alpha)\delta(0)\over (x_{12}\bar{x}_{12})^{2\Delta_{\alpha_1}}}.
\ea The $\delta(0)$ is proportional to the volume of the dilation group $\text{Vol({dilaton})}=\int_0^\infty {d\lambda\over \lambda}=\infty$.
Then using the "reflection relationship" \cite{McElgin:2007ak} $V_\alpha=S(\alpha)V_{Q-\alpha}$, one can obtain
\be
\label{twopointLiouville}\langle V_{\bar\alpha}(x_1)V_{\alpha}(x_2) \rangle_{\Sigma_1}={\delta(0) \over (x_{12}\bar{x}_{12})^{2\Delta_{\alpha_1}}}.
\ee

In terms of formula Eq.(\ref{ratio}) and the early time limit, the ratio becomes
\ba
R_{EE}^{(2)}& \underset{(z,\bar z)\to (0,0)}{\simeq}&\lim_{(z,\bar z)\to (0,0)}{\langle V_{\bar\alpha} V_{\alpha} V_{\bar\alpha} V_{\alpha} \rangle_{\Sigma_2}\over \langle V_{\bar\alpha} V_{\alpha}  \rangle_{\Sigma_1}^2} \underset{(z,\bar z)\to (0,0)}{ \simeq}  {\sqrt{\pi}\over 8\times 2! }{1\over \delta^2(0)}{d^2f_{\alpha}(p)\over d p^2}\Big|_{p\to 0} |z|^{2 \Delta_{{Q/2} }}\ln^{-{3\over 2}}|1/z|.\nonumber
\ea One can choose the appropriate normalisation condition to remove the $\delta^2(0)$ dependence\footnote{{Following standard regularisation from Eq.(5.13) in \cite{Nakayama:2004vk}, the normalisation factor can absorb $\delta^2(0)$. Here, we just keep the factor $\delta^2(0)$ like \cite{Harlow:2011ny} .}}.

Then
\ba\label{early-1}
 S_{EE}^{(2)} \underset{(z,\bar z)\to (0,0)}{\simeq}-\log(R_{EE}^{(2)}) \underset{(z,\bar z)\to (0,0)}{\simeq}-\log({\sqrt{\pi}\over 8\times 2! }\frac
 1 {\delta^2(0)}{d^2f_{\alpha}(p)\over d p^2}\Big|_{p\to 0} |z|^{2 \Delta_{{Q/2} }}\ln^{-{3\over 2}}|1/z|).
\ea
{Since the second REE is associated with $\langle {V}_{\bar\alpha}{V}_{\alpha}{V}_{\bar\alpha}{V}_{\alpha}\rangle$, the identity operator cannot contribute to the intermediate channels. When the external Liouville momentums of the four-point function are $\alpha_i \in \{\alpha|\alpha=Q/2+ i p,  p\in  \mathbb{R}\}\bigcup \{\alpha|Q/2>\text{Re}(\alpha)> {Q/4}\}\bigcup \{\alpha|{Q/4>\text{Re}(\alpha)> {0}}\}$, the primary operators will not fuse into the identity operator\footnote{{Generally speaking, if the $Re(\alpha_1+\alpha_2)<Q/2$, there are discrete term (\ref{discretetermLFT}) presented then the identity operator will contribute to the 4-point function as an intermediate channel operator. Since the second REE is associated with $\langle {V}_{\bar\alpha}(0){V}_{\alpha}(z){V}_{\bar\alpha}(1){V}_{\alpha}(\infty)\rangle$ and $Re(\alpha_1+\alpha_2)\geq Q/2$, the discrete terms (\ref{discreteterm}) will not contribute to the second REE.}}. Therefore, the vacuum block in the intermediate channel will not contribute to the four-point function in the second REE. {To restore the quasi-particle picture \cite{Nozaki:2014hna}\cite{He:2014mwa} and make well defined quantity to shown the memory effect of REE, we have to choose an appropriate reference state $V_{\alpha_{r}}|0\rangle$ which is not a vacuum state as given in (\ref{class}).} In this paper, we choose a reference state which lives in the same class of target states and we can define the difference of the second REE in the early time limit as follows}
\bea
 \Delta S_{EE}^{(2)}\big[V_\alpha|0\rangle, V_{\alpha_{r}}|0\rangle\big](t\to 0)&=&S_{EE}^{(2)}\big[V_\alpha|0\rangle\big](t\to 0) -S_{EE}^{(2)}\big[V_{\alpha_{r}}|0\rangle\big](t\to 0)\nonumber\\& =&-\log\Big({{ f_{\alpha}''(p)}\over {f_{\alpha_r}''(p)}}\Big) \Big|_{p\to 0},\nonumber\\ \text{ }\text{ } \alpha, \alpha_r \in\{\alpha|\alpha=Q/2+ i p,  p\in  \mathbb{R}\}&\bigcup& \{\alpha|Q/2>\text{Re}(\alpha)> {Q/4}\}\bigcup \{\alpha|{Q/4>\text{Re}(\alpha)> {0}}\}.\nonumber\\
\eea
Here one can see that the $\Delta S_{EE}^{(2)}\big[V_\alpha|0\rangle, V_{\alpha_{r}}|0\rangle\big](t\to 0)$ is finite. {Explicitly, one can choose $\alpha_r=\alpha$ and the $\Delta S_{EE}^{(2)}\big[V_\alpha|0\rangle, V_{\alpha_{r}}|0\rangle\big](t\to 0)$ will be vanishing which shows that the quasi-particle picture has been restored. For $\alpha_r\neq\alpha$, $\Delta S_{EE}^{(2)}\big[V_\alpha|0\rangle, V_{\alpha_{r}}|0\rangle\big](t\to 0)$ is finite with time evolution, which does not contradict the quasi-particle picture. } When $\alpha, \alpha_r$ do not stay in the same class given by (\ref{class}), the early and late time limits of $\Delta S_{EE}^{(2)}\big[V_\alpha|0\rangle, V_{\alpha_{r}}|0\rangle\big]$ cannot be finite due to different divergent powers of logarithmic divergence, e.g., (\ref{early-1}) and (\ref{lc})\footnote{In remain part of this paper, one can refer to the divergent piece of $S_{EE}^{(2)}\big[V_\alpha|0\rangle\big]$ in early and late time limits respectively.}.

Generally speaking, for the four-point function $\langle {V}_{\alpha_1}(0){V}_{\alpha_2}(z){V}_{\alpha_3}(1){V}_{\alpha_4}(\infty)\rangle_{\Sigma_1}$ with external legs $\alpha_{i}$ with $\text{Re}(\alpha_{i})\in (0, {Q/2})$, we have to consider the discrete terms' contributions which have been reviewed in the Appendix \ref{discrete} and Appendix \ref{appendix2}. In our setup $\langle {V}_{\bar\alpha}(0){V}_{\alpha}(z){V}_{\bar\alpha}(1){V}_{\alpha}(\infty)\rangle_{\Sigma_1}$\footnote{{Here we have chosen the $\alpha_1=\bar{\alpha}=Q-\alpha, \alpha_2=\alpha, \text{Re}(\alpha_{i})\in (0, {Q/2})$, therefore the discrete term will not be involved.}}, there is a marginal case:
\begin{itemize}

 \item $\text{Re}(\alpha) = {Q/4}, \text{Im}(\alpha)\neq 0$ (marginal case).

Since the factor $C(\bar\alpha, \alpha, \alpha_s)C( \bar\alpha, \alpha, Q-\alpha_s)$ in Eq.(\ref{4ptfunction}) does not vanish at $\alpha_s = Q/2$, we have
\ba
\label{marginal11}
&& \langle {V}_{\alpha}(0){V}_{\alpha}(z){V}_{\alpha}(1){V}_{\alpha}(\infty)\rangle_{\Sigma_1}\nonumber\\
& \underset{(z,\bar z)\to (0,0)}{\simeq}&  { C(\bar\alpha, \alpha, \frac{Q}{2})^2}\int_{\BR} |z|^{-2(2\Delta_{\alpha}-\Delta{2\alpha}) - 2p^2} d p  \nonumber\\& \underset{(z,\bar z)\to (0,0)}{\simeq}&   {\sqrt{\pi}\over2}C(\bar\alpha, \alpha, \frac{Q}{2})^2 |z|^{-2(2\Delta_{\alpha} - \Delta_{Q/2})} \ln^{-\frac12}|1/z|\,.
\ea
The ratio for the second REE in the early time limit reads
\[R_{EE}^{(2)} \underset{(z,\bar z)\to (0,0)}{\simeq}  {\sqrt{\pi}\over2}{1\over \delta^2 (0)}  C(\bar\alpha, \alpha, \frac{Q}{2})^2  |z|^{2 \Delta_{{Q/2} }}\ln^{-{1\over 2}}|1/z|. \]
Then
\ba\label{lc}
S_{EE}^{(2)}\big[V_\alpha|0\rangle\big](t\to 0)& \underset{(z,\bar z)\to (0,0)}{\simeq} &-\log\left({\sqrt{\pi}\over2}  \frac
 {1}{\delta^2(0)}C(\bar\alpha, \alpha, \frac{Q}{2})^2 |z|^{2 \Delta_{{Q/2} }}\ln^{-{1\over 2}}|1/z|\right).\nonumber\\
\ea
The second REE has divergent factor $\ln^{-{1\over 2}}|1/z|$ which can be canceled by choosing reference state $ V_{\alpha_{r}}|0\rangle, \alpha_r \in \{a\in \mathbb{C}|\text{Re}(a) = {Q/4},\text{Im}(a)\neq 0\}$ to obtain finite $\Delta S_{EE}^{(2)}\big[V_\alpha|0\rangle, V_{\alpha_{r}}|0\rangle\big]$.
In the early time limit, the difference of the second REE between target states and the reference state is
\bea
 &&\Delta S_{EE}^{(2)}\big[V_\alpha|0\rangle, V_{\alpha_{r}}|0\rangle\big](t\to 0)=S_{EE}^{(2)}\big[V_\alpha|0\rangle\big](t\to 0)-S_{EE}^{(2)}\big[V_{\alpha_{r}}|0\rangle\big](t\to 0) \nonumber\\&=&-\log\Big({{ f_{\alpha}(p)}\over {f_{\alpha_r}(p)}}\Big) \Big|_{p\to 0},\text{ }\text{ }\alpha, \alpha_r \in \{a\in \mathbb{C}|\text{Re}(a) = {Q/4},\text{Im}(a)\neq 0\}.
\eea
\end{itemize}

Now we will calculate the late time limit $(z,\bar z)\to (1,0)$ of the second REE. In this limit, applying bootstrap equation to holomorphic conformal blocks will be convenient when extracting correct late time behaviour. The four-point function of primary fields in LFT can be expressed by holomorphic $t$-channel conformal blocks as given in Appendix \ref{reviewLFT}
\ba
\langle V_{\bar\alpha} V_{\alpha} V_{\bar\alpha} V_{\alpha} \rangle_{\Sigma_1}
&=&{1\over 2}|z_{13}|^{-4\Delta_{\alpha}}|z_{24}|^{-4\Delta_{\alpha}}\int'_\BR {dp\over 2\pi} C(\bar\alpha,\alpha, {Q\over 2}+i p) C(\bar\alpha,\alpha,  {Q\over 2}-i p)\nonumber\\
&&F_{s}(\Delta_{i=\bar\alpha,\alpha,\bar\alpha,\alpha},\Delta_t,{\bar z})\int_\BS d\alpha_t F^L_{\xa_s \xa_t}\big[ {}_{\alpha}^{\bar\alpha} {}_{\bar\alpha}^{\alpha} \big]F_{t}(\Delta_{i=\bar\alpha,\alpha,\bar\alpha,\alpha},\Delta_t,{z}) \label{thirdterm-3}
\ea

The prime of integration over intermediate momentum $p$ stands for contour integration over reals with some additional so-called discrete terms' contributions.
The integral in $\xa_t$ is over $\BS = \frac Q 2 + i \BR^+$.  {$F^L_{\xa_s \xa_t}\big[ {}_{\xa}^{\bar\xa} {}_{\bar\xa}^{\xa} \big]$ is the fusion matrix associated with transformation from $s$-channel to $t$-channel and it has been revisited in Appendix \ref{fusionLFT}.} More precisely,
the $G_{\bar\alpha,\alpha,\bar\alpha,\alpha}(z,\bar z)$  can be expressed by holomorphic t-channel conformal block in terms of conformal bootstrap equation as follows:
\ba
G_{\bar\alpha,\alpha,\bar\alpha,\alpha}(z,\bar z)&=&\sum_{\alpha'_s\in D}\tilde{ D}_{\alpha'_s}(z,\bar z)+\int_{\BR^+ }{dp\over 2\pi} C(\bar\alpha,\alpha, {\frac Q 2}+i p)C(\bar\alpha,\alpha,  {\frac Q 2}-i p)\nonumber\\&&F_{s}(\Delta_{i=\bar\alpha,\alpha,\bar\alpha,\alpha},\Delta_t,{\bar z})\int_\BS d\alpha_t F^L_{\xa_s \xa_t}\big[ {}_{\xa}^{\bar\xa} {}_{\bar\xa}^{\xa} \big]F_{t}(\Delta_{i=\bar\alpha,\alpha,\bar\alpha,\alpha},\Delta_t,{ z})\label{discreteterm}
\ea
where $\tilde{D}$ is the finite set of discrete terms which have been reviewed in Appendix \ref{discrete} and Appendix \ref{appendix2}. The $D$ is the set of double poles induced by the factors $C(\bar\alpha,\alpha, \alpha_s) C(\bar\alpha,\alpha, \bar{\alpha_s})$ and $\tilde{ D}_{\alpha'_s}(z,\bar z)$\footnote{For external Liouville momentum $\alpha\in\{\alpha|\alpha=Q/2+ i p, p\in  \mathbb{R}\}\bigcup \{\alpha|Q/2>\text{Re}(\alpha)> {Q/4}\}\bigcup \{\alpha|{Q/4>\text{Re}(\alpha)> {0}}\}$, there are no discrete terms.} is given by the last line in Eq.(\ref{intpdis11}).

In the late time limit, the leading contributions to the REE will be as follows:
\ba
\langle V_{\bar\alpha} V_{\alpha} V_{\bar\alpha} V_{\alpha} \rangle_{\Sigma_1} &\underset{(z,\bar z)\to (1,0)}{\simeq}&
{1\over 2}|z_{13}|^{-4\Delta}|z_{24}|^{-4\Delta}\int'_{\BR }{dp\over 2\pi} C(\bar\alpha,\alpha, {Q\over 2}+i p) C(\bar\alpha,\alpha,  {Q\over 2}-i p)\nonumber\\
&&\bar{z}^{\Delta_p-2\Delta}\int_{\BS} d\alpha_t F^L_{\xa_s \xa_t}\big[ {}_{\xa}^{\bar\xa} {}_{\bar\xa}^{\xa} \big]({1-z})^{\Delta_t-2\Delta}\nonumber\\
&\underset{(z,\bar z)\to (1,0)}{\simeq}&{1\over 2}|z_{13}|^{-4\Delta}|z_{24}|^{-4\Delta}\int_{\BR }{dp\over 2\pi} C(\bar\alpha,\alpha, {Q\over 2}+i p) C(\bar\alpha,\alpha,  {Q\over 2}-i p)\nonumber\\
&&\bar{z}^{\Delta_p-2\Delta}\int_{\BS} d\alpha_t F^L_{\xa_s \xa_t}\big[ {}_{\xa}^{\bar\xa} {}_{\bar\xa}^{\xa} \big]({1-z})^{\Delta_t-2\Delta}+|z_{13}|^{-4\Delta}|z_{24}|^{-4\Delta}\sum_{\alpha'_s\in D}\tilde{ D}_{\alpha'_s}(z,\bar z)\quad\nonumber\\\label{thirdterm-2}
\ea
For external Liouville momentum $\alpha\in\{\alpha|\alpha=Q/2+ i p, p\in  \mathbb{R}\}\bigcup \{\alpha|Q/2>\text{Re}(\alpha)> {Q/4}\}\bigcup \{\alpha|{Q/4>\text{Re}(\alpha)> {0}}\}$ and $p\neq 0$, we take late time limit of Eq.(\ref{thirdterm-2}) and the Eq.(\ref{ratio}) will be
\ba\label{LFT-late-1}
\lim_{(z,\bar z) \to (1,0)}\langle V_{\bar\alpha} V_{\alpha} V_{\bar\alpha} V_{\alpha} \rangle_{\Sigma_1} & \underset{(z,\bar z)\to (1,0)}{\simeq}&
{\pi\over 64\times 2! \sqrt{\pi}}{d^2f_{\alpha}(p_s)\over d p_s^2}\Big|_{p_s\to 0} {2(s_b'({Q}))^2}{{F^L_{Q/2,Q/2}}\big[ {}_{\xa}^{\bar\xa} {}_{\bar\xa}^{\xa} \big]\over |s_b({Q})|^2} \nonumber\\&&(1-z)^{\Delta_{Q/2 }-2\Delta_{\alpha}} \bar{z}^{\Delta_{Q/2 }-2\Delta_{\alpha}} \ln^{-{3/2}}\left({1\over { (1-z)}}\right) \ln^{-{3/2}}\left({1\over { \bar z}}\right)\,.\nonumber\\
\ea
{We have used the late time limit $(z,\bar z) \to (1,0)$ and saddle point approximation to extract the leading contribution from relevant term $\alpha_s={Q/2},\alpha_t={Q/2}$.}

In this limit, the ratio becomes
\ba
R_{EE}^{(2)}& \underset{(z,\bar z)\to (1,0)}{\simeq}&\lim_{ {(z,\bar z)\to (1,0)}}{\langle V_{\bar\alpha} V_{\alpha} V_{\bar\alpha} V_{\alpha} \rangle_{\Sigma_2}\over \langle V_{\bar\alpha} V_{\alpha}  \rangle_{\Sigma_1}^2}\nonumber\\
& \underset{(z,\bar z)\to (1,0)}{\simeq} &{\pi\over 64\times 2! \sqrt{\pi}}{1\over  \delta^2(0)}{d^2f_{\alpha}(p_s)\over d p_s^2}\Big|_{p_s\to 0} {2(s_b'({Q}))^2}{{F^L_{Q/2,Q/2}}\big[ {}_{\xa}^{\bar\xa} {}_{\bar\xa}^{\xa} \big]\over |s_b({Q})|^2} \nonumber\\&&(1-z)^{\Delta_{Q/2 }} \bar{z}^{\Delta_{Q/2 }} \ln^{-{3/2}}\left({1\over { (1-z)}}\right) \ln^{-{3/2}}\left({1\over { \bar z}}\right).
\ea
We use same normalisation for two-point Green function given in Eq.(\ref{twopointLiouville}).

Then the second REE in the late time limit reads
\bea
&&S_{EE}^{(2)}\big[V_\alpha|0\rangle\big](t\to \infty)\nonumber\\& \underset{(z,\bar z)\to (1,0)}{\simeq}&-\log\Big({\pi\over 64\times 2! \sqrt{\pi}}\frac
 1 {\delta^2(0)}{d^2f_{\alpha}(p)\over d p^2}\Big|_{p\to 0}{2(s_b'({Q}))^2}{{F^L_{Q/2,Q/2}}\big[ {}_{\xa}^{\bar\xa} {}_{\bar\xa}^{\xa} \big]\over |s_b({Q})|^2}\nonumber\\&& (1-z)^{\Delta_{Q/2 }} \bar{z}^{\Delta_{Q/2 }} \ln^{-{3/2}}\left({1\over { (1-z)}}\right) \ln^{-{3/2}}\left({1\over { \bar z}}\right)\Big).\label{integralsuper}
\eea
In the late time limit, the difference of the second REE between the target state and the reference state is
\ba
 \Delta S_{EE}^{(2)}\big[V_\alpha|0\rangle, V_{\alpha_{r}}|0\rangle\big](t\to \infty)&=&S_{EE}^{(2)}\big[V_\alpha|0\rangle\big](t\to \infty)-S_{EE}^{(2)}\big[V_{\alpha_{r}}|0\rangle\big](t\to \infty)\nonumber\\&=&-\log\Big({{ f_{\alpha}''(p)F^L_{{Q/2,Q/2}}\big[ {}_{\xa}^{\bar\xa} {}_{\bar\xa}^{\xa} \big]}\over {f_{\alpha_r}''(p)F^L_{{Q/2,Q/2}}\big[ {}_{\xa_r}^{\bar{\xa_r}} {}_{\bar{\xa_r}}^{{\xa_r}} \big]}}\Big) \Big|_{p\to 0},\nonumber\\ \text{ }\text{ } \alpha, \alpha_r \in \{\alpha|\alpha=Q/2+ i p,  p\in  \mathbb{R}\}&\bigcup& \{\alpha|Q/2>\text{Re}(\alpha)> {Q/4}\}\bigcup \{\alpha|{Q/4>\text{Re}(\alpha)> {0}}\}.\nonumber\\
\ea

Additionally, we have to consider the marginal case:
\begin{itemize}

\item $\text{Re}(\alpha) = {Q/4}, \text{Im}(\alpha)\neq 0$ (marginal case).

Since $C( \bar\alpha, \alpha, \alpha_s)C(\bar\alpha, \alpha, Q-\alpha_s)$ does not vanish at $\alpha_s = Q/2$, the late time of the 4-point function behaves
\begin{align}
\label{LFT-late-4}
&\langle {V}_{\bar\alpha}(0){V}_{\alpha}(z){V}_{\bar\alpha}(1){V}_{\alpha}(\infty)\rangle_{\Sigma_1}\nonumber\\
 \underset{(z,\bar z)\to (1,0)}{\simeq}& C(\bar\alpha, \alpha, \frac{Q}{2})^2 \int_{\BR} F^L_{\alpha_s=Q/2+ip_s,\alpha_t=Q/2+i p_t}\big[ {}_{\xa}^{\xa} {}_{\xa}^{\xa} \big](1-z)^{\Delta_{2\alpha}-2\Delta_\alpha + p_s^2}(\bar z)^{\Delta_{2\alpha}-2\Delta_\alpha + p_t^2} d p_s d p_t\nonumber\\
 \underset{(z,\bar z)\to (1,0)}{\simeq}&  {\pi\over 32} C(\bar\alpha, \alpha, \frac{Q}{2})^2 {2(s_b'({Q/2}))^2}{{F^L_{Q/2,Q/2}}\big[ {}_{\xa}^{\bar\xa} {}_{\bar\xa}^{\xa} \big]\over |s_b({Q/2})|^2} \nonumber\\&(1-z)^{\Delta_{Q/2}-2\Delta_\alpha }(\bar z)^{\Delta_{Q/2}-2\Delta_\alpha } \ln^{-{1\over 2}}\left({1\over { (1-z)}}\right)\ln^{-{3\over 2}}\left({1\over { \bar z}}\right)\,.
\end{align}

In the late time limit, the ratio for the second REE is
\ba R_{EE}^{(2)}& \underset{(z,\bar z)\to (1,0)}{\simeq}&   {\pi\over 32}\frac
 1 {\delta^2(0)}C(\bar\alpha, \alpha, \frac{Q}{2})^2 {2(s_b'({Q/2}))^2}{{F^L_{Q/2,Q/2}}\big[ {}_{\xa}^{\bar\xa} {}_{\bar\xa}^{\xa} \big]\over |s_b({Q/2})|^2} (1-z)^{\Delta_{Q/2} }(\bar z)^{\Delta_{Q/2} }\nonumber\\&& \ln^{-{1\over 2}}\left({1\over { (1-z)}}\right)\ln^{-{3\over 2}}\left({1\over { \bar z}}\right)\,.\ea
{Then}
\ba
&&S_{EE}^{(2)}\big[V_\alpha|0\rangle\big](t\to \infty)\nonumber\\& \underset{(z,\bar z)\to (1,0)}{\simeq}&  -\log \Big[  {\pi\over 32} \frac {1 }{\delta^2(0)}C(\bar\alpha, \alpha, \frac{Q}{2})^2 {2(s_b'({Q/2}))^2}{{F^L_{Q/2,Q/2}}\big[ {}_{\xa}^{\bar\xa} {}_{\bar\xa}^{\xa} \big]\over |s_b({Q/2})|^2} \nonumber\\&&(1-z)^{\Delta_{Q/2 }}  \bar{z}^{\Delta_{Q/2 }}\ln^{-{1\over 2}}\left({1\over { (1-z)}}\right)\ln^{-{3\over 2}}\left({1\over { \bar z}}\right)\Big],
\ea
and the difference of the second REE between the target state and the reference state in the late time limit
is
\bea
&&\Delta S_{EE}^{(2)}\big[V_\alpha|0\rangle, V_{\alpha_{r}}|0\rangle\big](t\to \infty)=S_{EE}^{(2)}\big[V_\alpha|0\rangle\big](t\to \infty)-S_{EE}^{(2)}\big[V_{\alpha_{r}}|0\rangle\big](t\to \infty)\nonumber\\& =&-\log\Big({{ f_{\alpha}(p)}F^L_{{{Q/2}},{Q/2}}\big[ {}_{\xa}^{\bar\xa} {}_{\bar{\xa}}^{ \xa} \big]\over {f_{\alpha_r}(p)}F^L_{{{Q/2}},{Q/2}}\big[ {}_{\xa_r}^{\bar{\xa}_r} {}_{\bar{\xa}_r}^{{\xa_r}} \big]}\Big) \Big|_{p\to 0}\text{ }\text{ }\alpha, \alpha_r \in \{a\in \mathbb{C}|\text{Re}(a) = {Q/4},\text{Im}(a)\neq 0\}.\nonumber\\
\eea
\end{itemize}

\subsection{The second REE in Super Liouville field theory}
{In this section}, we would like to consider the states excited by local operators in super Liouville field theory. For sake of consistent notation, we review the corresponding contents of SLFT in Appendix \ref{SLFT}.
The four-point Green function for NS-NS operator $V_{\bar\alpha}, V_{\alpha} $ with s-channel intermediate states Eq.(\ref{4pt-SLFT}) is as follows,
\ba
&&\langle V_{\bar\alpha} V_{\alpha} V_{\bar\alpha} V_{\alpha} \rangle_{\Sigma_1} \nonumber\\&=&|z_{13}|^{-4\Delta}|z_{24}|^{-4\Delta} G_{\bar{1}2\bar{3}4}({z,\bar{z}})\nonumber\\
&=&|z_{13}|^{-4\Delta}|z_{24}|^{-4\Delta}\Big[\int'_{S}{d\alpha_s} C_{NS}(\bar\alpha,\alpha, \alpha_s) C_{NS}(\bar\alpha,\alpha,  \bar\alpha_s)F_{s}^{e}(\Delta_{i=1,2,3,4},\Delta_p,\bar{z}) F_{s}^{e}(\Delta_{i=1,2,3,4},\Delta_p,{z})\nonumber\\
&+& \int'_{S}{d\alpha_s} \tilde{C}_{NS}(\bar\alpha,\alpha, \alpha_s) \tilde{C}_{NS}(\bar\alpha,\alpha, \alpha_s)F_{s}^{o}(\Delta_{i=1,2,3,4},\Delta_p,\bar{z})F_{s}^{o}(\Delta_{i=1,2,3,4},\Delta_p,{z})\Big]\label{thirdterm-21}
\ea
The four point green function for R-R operator $R_{\bar\alpha}, R_{\alpha}$ reads similarly,
\ba
&&\langle R_{\bar\alpha} R_{\alpha} R_{\bar\alpha} R_{\alpha} \rangle_{\Sigma_1}\nonumber\\ &=&|z_{13}|^{-4\Delta}|z_{24}|^{-4\Delta} G_{\bar{1}2\bar{3}4}({z,\bar{z}})\nonumber\\
&=&|z_{13}|^{-4\Delta}|z_{24}|^{-4\Delta}\Big[\int'_{S}{d\alpha_s} C_R(\bar\alpha,\alpha, \alpha_s) C_R(\bar\alpha,\alpha, \bar\alpha_s)F_{s}^{e}(\Delta_{i=\bar\alpha,\alpha,\bar\alpha,\alpha},\Delta_p,\bar{z}) F_{s}^{e}(\Delta_{i=\bar\alpha,\alpha,\bar\alpha,\alpha},\Delta_p,{z})\nonumber\\
&+& \int'_{S}{d\alpha_s} \tilde{C}_R(\bar\alpha,\alpha, \alpha_s) \tilde{C}_R(\bar\alpha,\alpha, \alpha_s)F_{s}^{o}(\Delta_{i=\bar\alpha,\alpha,\bar\alpha,\alpha},\Delta_p,\bar{z})F_{s}^{o}(\Delta_{i=\bar\alpha,\alpha,\bar\alpha,\alpha},\Delta_p,{z})\Big]\label{RR4point}
\ea
All the calculations of REE for the NS-NS states can be directly generalised to states excited by the R-R operators and hence we only carry out the analysis in the former case (\ref{thirdterm-21}).

{We start with external super Liouville momentum $\xa_i = \{\alpha|\alpha=Q/2+ i p, p\in  \mathbb{R}\}\bigcup \{\alpha|Q/2>\text{Re}(\alpha)> {Q/4}\}\bigcup \{\alpha|{Q/4>\text{Re}(\alpha)> {0}}\}$.} With saddle point approximation, the early time behaviour of Eq.(\ref{thirdterm-21}) is as follows:
\ba\label{subleadingslft}
\lim_{(z,\bar z) \to (0,0)}\langle V_{\bar\alpha} V_{\alpha} V_{\bar\alpha} V_{\alpha} \rangle_{\Sigma_1} &\underset{(z,\bar z)\to (0,0)}{\simeq}& {1\over 2!}{d^2{f_{\alpha}}_{NS}(p)\over d p^2}\Big|_{p\to 0}  \int_{\mathbb{R}} |z|^{-2(2\Delta_{\alpha}  - \Delta_{Q/2}) +2p^2} p^2 \, d p \nonumber\\& \underset{(z,\bar z)\to (0,0)}{\simeq}& {\sqrt{\pi}\over 4\times 2!}{d^2{f_{\alpha}}_{NS}(p)\over d p^2}\Big|_{p\to 0} |z|^{-2(2\Delta_{\alpha}  - \Delta_{Q/2})} \ln^{-\frac32}|1/z|\,,\nonumber\\
\ea
where ${f_{\alpha}}_{NS}(p_s)=C_{NS}(\bar\alpha,\alpha, \alpha_s) C_{NS}(\bar\alpha,\alpha,  \bar\alpha_s)$ and $\tilde{f_{\alpha}}_{NS}=\tilde{C}_{NS}(\bar\alpha,\alpha, \alpha_s) \tilde{C}_{NS}(\bar\alpha,\alpha, \alpha_s)$. {In the early time limit $(z,\bar z)\to(0,0)$, we have used the fact that the leading intermediate state in the parity odd conformal block $F_{s}^{o}(\Delta_{i=\bar{\alpha},\alpha,\bar{\alpha},\alpha},\Delta_p,\bar{z})F_{s}^{o}(\Delta_{i=\bar{\alpha},\alpha,\bar{\alpha},\alpha},\Delta_p,{z})$ in (\ref{thirdterm-21}) is $G_{-1/2}\tilde G_{-1/2}V_{\xa_s}$ and hence its contribution will be smaller by a factor of $z^{1/2}$ compared to the even conformal block (see \cite{Hadasz:2008dt} for more details). As a result we can make the contribution from the parity-even conformal block \cite{Poghosyan:2016kvd}.}
For R-R sector, we directly replace $C_{NS}({\alpha_{1},\alpha_{2},\alpha_{3}})$ with $C_R({\alpha_{1},\alpha_{2},\alpha_{3}})$. The structure constant $C_R({\alpha_{1},\alpha_{2},\alpha_{s}})$ has the simple pole at $\xa_s = Q/2$  as $C_{NS}({\alpha_{1},\alpha_{2},\alpha_{s}})$, which is from $\Upsilon_{\zt{NS}}(\xa_3)$ in the numerator of \er{3ptfunctionSLFT} and \er{dozzRR}, therefore, the analysis of the local excited states associated with R-R operator is the same as those with an NS-NS operator.

The two-point Green function for primary operator in NS sector is as follows:
%\ba
%\langle V_{\alpha}(x_1)V_{\alpha}(x_2) \rangle={2\pi D_{NS}(\alpha)\over (x_{12}\bar{x}_{12})^{2\Delta_{\alpha_1}}}(1+2 \Delta {\theta_1\theta_2\over x_{12}}+2 \Delta %{\bar\theta_1\bar\theta_2\over \bar x_{12}}+ {\theta_1\theta_2 \bar\theta_1\bar\theta_2\over x_{12}\bar x_{12}})
%\ea
\ba\label{SLFT-late-1}
\langle V_{\alpha}(x_1)V_{\alpha}(x_2) \rangle_{\Sigma_1}= { D_{NS}(\alpha)\delta(0)\over (x_{12}\bar{x}_{12})^{2\Delta_{\alpha}}}
\ea
with
\ba
D_{NS}(\alpha)=(\pi \mu \gamma({bQ\over 2}))^{(Q-2\alpha)\over b}{b^2\gamma\left( b\alpha-{1\over 2}-{b^2\over 2}\right)\over \gamma\left({1\over 2}+{b^{-2}\over 2}-\alpha b^{-1}\right)}.
\ea
Then using the ``reflection relationship" $V_\alpha=D_{NS}(\alpha)V_{Q-\alpha}$, one can obtain
\ba
\langle V_{\bar\alpha}(x_1)V_{\alpha}(x_2) \rangle_{\Sigma_1}= {\delta(0)\over (x_{12}\bar{x}_{12})^{2\Delta_{\alpha}}}\,.
\ea
{The associated ratio in the early time limit is
\ba
R_{EE}^{(2)}\underset{(z,\bar z)\to (0,0)}{\simeq}{\langle V_{\bar\alpha} V_{\alpha} V_{\bar\alpha} V_{\alpha} \rangle_{\Sigma_2}\over \langle V_{\bar\alpha} V_{\alpha}  \rangle_{\Sigma_1}^2}&\underset{(z,\bar z)\to (0,0)}{\simeq}&  {\sqrt{\pi}\over 8\times 2!}{1\over  \delta^2(0)} {d^2{f_{\alpha}}_{NS}(p)\over d p^2}\Big|_{p\to 0}|z|^{2 \Delta_{{Q/2} }}\ln^{-{3\over 2}}|1/z|\nonumber
\ea As we have done in LFT, we also keep the normalisation factor with a delta function in explicit form.}

Then
\ba
S_{EE}^{(2)}(t\to 0)=-\log(R_{EE}^{(2)})&\underset{(z,\bar z)\to (0,0)}{\simeq} & -\log({\sqrt{\pi}\over 8\times 2!}\frac
 1 {\delta^2(0)}{d^2{f_{\alpha}}_{NS}(p)\over d p^2}\Big|_{p\to 0}|z|^{2 \Delta_{{Q/2} }}\ln^{-{3\over 2}}|1/z|).\nonumber
\ea
Finally, the early time of the difference of second REE between $V_\alpha|0\rangle$ and $ V_{\alpha_{r}}|0\rangle$ is
\ba
\Delta S_{EE}^{(2)}\big[V_\alpha|0\rangle, V_{\alpha_{r}}|0\rangle\big](t\to 0)&=&S_{EE}^{(2)}\big[V_\alpha|0\rangle\big](t\to 0)-S_{EE}^{(2)}\big[V_{\alpha_{r}}|0\rangle\big](t\to 0) \nonumber\\&=&-\log\Big({{ {f_{\alpha}}_{NS}''(p)}\over {{f_{\alpha_r}}_{NS}''(p)}}\Big) \Big|_{p\to 0}, \nonumber\\\text{ }\text{ } \alpha, \alpha_r \in \{\alpha|\alpha=Q/2+ i p,  p\in  \mathbb{R}\}&\bigcup& \{\alpha|Q/2>\text{Re}(\alpha)> {Q/4}\}\bigcup \{\alpha|{Q/4>\text{Re}(\alpha)> {0}}\} .\nonumber\\
\ea
Additionally, we have to consider the marginal case:
\begin{itemize}

\item $\text{Re}(\alpha) = \frac{Q}{4}, \text{Im}(\alpha)\neq 0$ (marginal case).

This case is similar to that mentioned above, except that $C_{NS}(\bar\alpha, \alpha, \alpha_s,)C_{NS}( \bar\alpha, \alpha, Q-a_s)$ does not vanish at $\alpha_s = Q/2$, so we have
\ba
\label{SLFT-late-4}
&& \langle {V}_{\bar\alpha}(0){V}_{\alpha}(z){V}_{\bar\alpha}(1){V}_{\alpha}(\infty)\rangle_{\Sigma_1}\nonumber\\&
\underset{(z,\bar z)\to (0,0)}{\simeq} & C_{NS}(\bar\alpha, \alpha, \frac{Q}{2})^2\int_{R} |z|^{-2(2\Delta_{2\alpha}-\Delta_{\alpha}) + 2p^2} d p  \nonumber\\&\underset{(z,\bar z)\to (0,0)}{\simeq}& {\sqrt{\pi}\over 2} C_{NS}(\bar\alpha, \alpha, \frac{Q}{2})^2 |z|^{-2(\Delta_{\alpha}  - \Delta_{Q/2})} \ln^{-\frac12}|1/z|\,.
\ea

{The ratio for the second REE in the early time limit is}
\ba \label{sc}R_{EE}^{(2)}\underset{(z,\bar z)\to (0,0)}{\simeq} {\sqrt{\pi}\over 2}{1\over \delta^2(0) } C_{NS}(\bar\alpha, \alpha, \frac{Q}{2})^2 |z|^{2 \Delta_{{Q/2} }}\ln^{-{1\over 2}}|1/z|. \ea
Then
\ba
 S_{EE}^{(2)}\big[V_\alpha|0\rangle\big](t\to 0)&\underset{(z,\bar z)\to (0,0)}{\simeq} & -\log\left({\sqrt{\pi}\over 2}{1\over \delta^2(0) } C_{NS}(\bar\alpha, \alpha, \frac{Q}{2})^2 |z|^{2 \Delta_{{Q/2} }}\ln^{-{1\over 2}}|1/z|\right).\nonumber\\
\ea
The second REE has divergent factor $\ln^{-{1\over 2}}|1/z|$ which can be canceled by choosing reference state $ V_{\alpha_{r}}|0\rangle,\alpha_r \in \{{a \in \mathbb{C}}|\text{Re}(a) = {Q/4}, \text{Im}(a)\in  \mathbb{R}, \text{Im}(\alpha)\neq 0\}$.
Finally, the early time of the difference of the second REE between $V_\alpha|0\rangle$ and $ V_{\alpha_{r}}|0\rangle$ is
\bea
 &&\Delta S_{EE}^{(2)}\big[V_\alpha|0\rangle, V_{\alpha_{r}}|0\rangle\big](t\to 0)=S_{EE}^{(2)}\big[V_\alpha|0\rangle\big](t\to 0)-S_{EE}^{(2)}\big[V_{\alpha_{r}}|0\rangle\big](t\to 0) \nonumber\\& =&-\log\Big({{ C_{NS}(\bar\alpha, \alpha, \frac{Q}{2})^2}\over {C_{NS}(\bar{\alpha_r}, \alpha_r, \frac{Q}{2})^2}}\Big)\nonumber\\&& \text{ }\text{ }\alpha, \alpha_r \in \{{a \in \mathbb{C}}|\text{Re}(a) = {Q/4}, \text{Im}(a)\in  \mathbb{R}, \text{Im}(\alpha)\neq 0\}.
\eea
\end{itemize}

Secondly, we consider the second REE in SLFT in the late time limit. For convenience in the late time limit, we have to use conformal bootstrap equation to express the four-point function $G_{1234}({z,\bar{z}})$, which is similar to the procedures shown in LFT. The four-point function can be expressed as follows:
\ba
&&\langle V_{\alpha_1}(0,0) V_{\alpha_2}(z,\bar z) V_{\alpha_3}(1,1) V_{\alpha_4}(\infty, \infty)\rangle_{\Sigma_1}= G_{1234}({z,\bar{z}})\nonumber\\
&=&{1\over 2}\Big(\int'_{S}{d\alpha_s} C_{NS}(\alpha_1,\alpha_2, \alpha_s) C_{NS}(\alpha_3,\alpha_4,  \bar \alpha_s)\nonumber\\
&&F_{s}^{e}(\Delta_{i=1,2,3,4},\Delta_{\alpha_s},\bar{z})\int d\alpha_t \sum_{\rho=e,o}F^{SL}_{\xa_s \xa_t}\big[ {}_{\xa}^{\bar\xa} {}_{\bar\xa}^{\xa} \big]^{e}_{\rho}F_{t}^{\rho}(\Delta_{i=1,2,3,4},\Delta_{\alpha_t},{z})\nonumber\\
&+& \int'_{S}{d\alpha_s} \tilde{C}_{NS}(\alpha_1,\alpha_2, \alpha_s) \tilde{C}_{NS}(\alpha_3,\alpha_4,  \bar \alpha_s)F_{s}^{o}(\Delta_{i=1,2,3,4},\Delta_{\alpha_s},\bar{z})\nonumber\\&&\int d\alpha_t \sum_{\rho=e,o} F^{SL}_{\xa_s \xa_t}\big[ {}_{\xa_4}^{\xa_3} {}_{\xa_1}^{\xa_2} \big]^{o}_{\rho}F_{t}^{\rho}(\Delta_{i=1,2,3,4},\Delta_{\alpha_t},{z})\Big).\label{thirdterm-22}
\ea
In the late time limit of Eq.(\ref{thirdterm-22}) with saddle point approximation, the four point
function becomes
\ba\label{SLFT-late-1L}
&&\lim_{(z,\bar z) \to (1,0)}\langle V_{\bar\alpha} V_{\alpha} V_{\bar\alpha} V_{\alpha} \rangle_{\Sigma_1} \nonumber\\&\underset{(z,\bar z)\to (1,0)}{\simeq} &{1\over  2!}{d^2{f_{\alpha}}_{NS}(p_s)\over d p_s^2}\Big|_{p_s\to 0}\nonumber\\&&\int_{\mathbb{R}}F^{SL}_{\alpha_s=Q / 2+ i p_s,\alpha_t=Q/2+i p_t}\big[ {}_{\xa}^{\bar\xa} {}_{\bar\xa}^{\xa} \big]^{e}_{e}\left(1-z\right)^{\Delta_{Q/2 }-2\Delta_{\alpha}+ p^2} (\bar z)^{\Delta_{Q/2 }-2\Delta_{\alpha} + p_s^2} p_s^2 \, d p_s d p_t .\nonumber\\
&\underset{(z,\bar z)\to (1,0)}{\simeq}& {{\pi}\over 64\times 2!}{d^2{f_{\alpha}}_{NS}(p_s)\over d p_s^2}\Big|_{p_s\to 0} {2(s_{NS}'({Q}))^2}{F^{SL}_{{{Q/2}},{Q/2}}\big[ {}_{\xa}^{\bar\xa} {}_{\bar\xa}^{\xa} \big]^{e}_{e}\over |s_{NS}({Q})|^2} \nonumber\\&&(1-z)^{\Delta_{Q/2 }-2\Delta_{\alpha}} \bar{z}^{\Delta_{Q/2 }-2\Delta_{\alpha}} \ln^{-{3/2}}\left({1\over { (1-z)}}\right) \ln^{-{3/2}}\left({1\over { \bar z}}\right).
\ea
where the exact expression for ${F^{SL}_{{{Q/2}},{Q/2}}\big[ {}_{\xa}^{\bar\xa} {}_{\bar\xa}^{\xa} \big]}$ is revisited in Appendix \ref{fusionSLFT}.
We have used the late time limit to extract the leading contribution in the final step of Eq.(\ref{SLFT-late-1L}). {One can show that the contribution from odd parity part of conformal block $F_{t}^{o}(\Delta_{i=1,2,3,4},\Delta_p,{z})$ in late time limit will be sub-leading \cite{Poghosyan:2016kvd} and we drop the sub-leading contributions in Eq.(\ref{SLFT-late-1L}).}

Then the ratio associated with the second REE in super Liouville field theory can be defined as follows:
\ba
R_{EE}^{(2)}&=& \lim_{z\to 1,\bar{z}\to 0}{\langle V_{\bar\alpha} V_{\alpha} V_{\bar\alpha} V_{\alpha} \rangle_{\Sigma_2}\over \langle V_{\bar\alpha} V_{\alpha}  \rangle_{\Sigma_1}^2}\nonumber\\ &\underset{(z,\bar z)\to (1,0)}{\simeq} & {{\pi}\over 64\times 2!}{d^2{f_{\alpha}}_{NS}(p_s)\over d p_s^2}\Big|_{p_s\to 0}{1\over  \delta^2(0)} {2(s_{NS}'({Q}))^2} {F^{SL}_{{{Q/2}},{Q/2}}\big[ {}_{\xa}^{\bar\xa} {}_{\bar\xa}^{\xa} \big]^{e}_{e}\over |s_{NS}({Q})|^2} \nonumber\\&&(1-z)^{\Delta_{Q/2 }} \bar{z}^{\Delta_{Q/2 }} \ln^{-{3/2}}\left({1\over { (1-z)}}\right) \ln^{-{3/2}}\left({1\over { \bar z}}\right).
\ea
Then the corresponding the second REE in the late time limit is
\ba
&&S_{EE}^{(2)}\big[V_\alpha|0\rangle\big](t\to \infty)\nonumber\\&\underset{(z,\bar z)\to (1,0)}{\simeq}&-\log \Big({{\pi}\over 64\times 2!}\frac
 1 {\delta^2(0)}{d^2{f_{\alpha}}_{NS}(p_s)\over d p_s^2}\Big|_{p_s\to 0}{2(s_{NS}'({Q}))^2} {F^{SL}_{{{Q/2}},{Q/2}}\big[ {}_{\xa}^{\bar\xa} {}_{\bar\xa}^{\xa} \big]^{e}_{e}\over |s_{NS}({Q})|^2} \nonumber\\&&(1-z)^{\Delta_{Q/2 }} \bar{z}^{\Delta_{Q/2 }} \ln^{-{3/2}}\left({1\over { (1-z)}}\right) \ln^{-{3/2}}\left({1\over { \bar z}}\right) \Big).\label{finalREE}
\ea
Choosing the same corresponding reference state as we used in the early time situation, the late time of the difference $\Delta S_{EE}^{(2)}\big[V_\alpha|0\rangle, V_{\alpha_{r}}|0\rangle\big]$ is
\ba
\Delta S_{EE}^{(2)}\big[V_\alpha|0\rangle, V_{\alpha_{r}}|0\rangle\big](t\to \infty)&=&S_{EE}^{(2)}\big[V_\alpha|0\rangle\big](t\to \infty)-S_{EE}^{(2)}\big[V_{\alpha_{r}}|0\rangle\big](t\to \infty) ,\nonumber\\&=&-\log\Big({{ {f_{\alpha}}_{NS}''(p)F^{SL}_{{Q/2,Q/2}}\big[ {}_{\xa}^{\bar\xa} {}_{\bar\xa}^{\xa} \big]}\over {{f_{\alpha_r}}_{NS}''(p)F^{SL}_{{Q/2,Q/2}}\big[ {}_{\xa_r}^{\bar{\xa_r}} {}_{\bar{\xa_r}}^{{\xa_r}} \big]}}\Big) \Big|_{p\to 0},\nonumber\\\text{ }\text{ } \alpha, \alpha_r \in \{\alpha|\alpha=Q/2+ i p,  p\in  \mathbb{R}\}&\bigcup& \{\alpha|Q/2>\text{Re}(\alpha)> {Q/4}\}\bigcup \{\alpha|{Q/4>\text{Re}(\alpha)> {0}}\}.\nonumber\\
\ea
Additionally, we have to consider the marginal case:
\begin{itemize}

\item $\text{Re}(\alpha) = \frac{Q}{4}, \text{Im}(\alpha)\neq 0$ (marginal case).

This case is similar to that mentioned above, except that $ C_{NS}(\bar\alpha, \alpha, \frac{Q}{2})^2$ does not vanish at $\alpha_s = Q/2$, so we have
\begin{align}
\label{SLFT-late-4L}
&\langle {V}_{\bar\alpha}(0){V}_{\alpha}(z){V}_{\bar\alpha}(1){V}_{\alpha}(\infty)\rangle_{\Sigma_1}\nonumber\\
\underset{(z,\bar z)\to (1,0)}{\simeq}&  C_{NS}(\bar\alpha, \alpha, \frac{Q}{2})^2\int_{R} F^{SL}_{\alpha_s=Q/2+i p_s,\alpha_t=Q/2+i p_t}\big[ {}_{\xa}^{\xa} {}_{\xa}^{\xa} \big]^{e}_{e}(1-z)^{\Delta_{2\alpha}-2\Delta_\alpha + p_s^2}(\bar z)^{\Delta_{2\alpha}-2\Delta_\alpha + p_t^2} d p_s d p_t\nonumber\\
\underset{(z,\bar z)\to (1,0)}{\simeq}& {\pi\over 16\times 2} C_{NS}(\bar\alpha, \alpha, \frac{Q}{2})^2 {2(s_{NS}'(Q))^2} {F^{SL}_{{{Q/2}},{Q/2}}\big[ {}_{\xa}^{\xa} {}_{\xa}^{\xa} \big]^{e}_{e}\over s_{NS}(Q)^2 }\nonumber\\&(1-z)^{\Delta_{2\alpha}-2\Delta_\alpha }(\bar z)^{\Delta_{2\alpha}-2\Delta_\alpha }\ln^{-{1\over 2}}\left({1\over { (1-z)}}\right)\ln^{-{3\over 2}}\left({1\over { \bar z}}\right)\,.
\end{align}

For $\text{Re}(\alpha)={Q\over 4}$, in the late time limit, the ratio for the second REE reads
\ba R_{EE}^{(2)}&\underset{(z,\bar z)\to (1,0)}{\simeq}& {\pi\over 16\times 2}{1\over  \delta^2(0) } C_{NS}(\bar\alpha, \alpha, \frac{Q}{2})^2 {2(s_{NS}'(Q))^2} {F^{SL}_{{{Q/2}},{Q/2}}\big[ {}_{\xa}^{\xa} {}_{\xa}^{\xa} \big]^{e}_{e}\over s_{NS}(Q)^2 }\nonumber\\&&(1-z)^{\Delta_{Q/2 }}  \bar{z}^{\Delta_{Q/2 }}\ln^{-{1\over 2}}\left({1\over { (1-z)}}\right)\ln^{-{3\over 2}}\left({1\over { \bar z}}\right).\ea
{Then}
\bea
 &&S_{EE}^{(2)}\big[V_\alpha|0\rangle\big](t\to \infty)\nonumber\\&\underset{(z,\bar z)\to (1,0)}{\simeq} & -\log \Big[ {\pi\over 16\times 2}\frac{
 1}{\delta^2(0)}C_{NS}(\bar\alpha, \alpha, \frac{Q}{2})^2  {2(s_{NS}'(Q))^2} {F^{SL}_{{{Q/2}},{Q/2}}\big[ {}_{\xa}^{\xa} {}_{\xa}^{\xa} \big]^{e}_{e}\over s_{NS}(Q)^2 }\nonumber\\&&(1-z)^{\Delta_{Q/2 }}  \bar{z}^{\Delta_{Q/2 }}\ln^{-{1\over 2}}\left({1\over { (1-z)}}\right)\ln^{-{3\over 2}}\left({1\over { \bar z}}\right)\Big].
\eea
Finally
\bea
 &&\Delta S_{EE}^{(2)}\big[V_\alpha|0\rangle, V_{\alpha_{r}}|0\rangle\big](t\to \infty)=S_{EE}^{(2)}\big[V_\alpha|0\rangle\big](t\to \infty)-S_{EE}^{(2)}\big[V_{\alpha_{r}}|0\rangle\big](t\to \infty)\nonumber\\& =&-\log\Big({{ {f_{\alpha}}_{NS}(p)}F^{SL}_{{{Q/2}},{Q/2}}\big[ {}_{\xa}^{\bar\xa} {}_{\bar{\xa}}^{ \xa} \big]\over {{f_{\alpha_r}}_{NS}(p)}F^{SL}_{{{Q/2}},{Q/2}}\big[ {}_{\xa_r}^{\bar{\xa}_r} {}_{\bar{\xa}_r}^{{\xa_r}} \big]}\Big) \Big|_{p\to 0}\nonumber\\&&\text{ }\text{ }\alpha, \alpha_r \in \{a \in {\mathbb{C}}|\text{Re}(a) = {Q/4}, \text{Im}(a)\in  \mathbb{R}, \text{Im}(\alpha)\neq 0\}.
\eea

\end{itemize}
One can show that we have to choose an appropriate reference states in terms of classification (\ref{class}) in LFT and SLFT. {Then the early time and late time behavior of $\Delta S_{EE}^{(2)}\big[V_\alpha|0\rangle, V_{\alpha_{r}}|0\rangle\big](t)$ can be well defined and the quasi-particle picture can be restored.} That also means that the divergent behaviour of REE of local excited states in LFT and SLFT is quite different from that in the vacuum state, although the vacuum state is excluded in LFT and SLFT as argued in \cite{vacuumstate}. Our calculations have classified the divergent behaviours of entanglement entropy of local excited states, especially in LFT and SLFT.
\section{The $n$th REE in LFT and SLFT}
{In the previous section, we computed the second REE of the local excited states.
In this sub-section, we use the n-point conformal block and operator product expansion (OPE) to obtain the $n$th R\'enyi entanglement entropy of the local excited states.}
Here we give a sketch of the $n$th REE with following \cite{He:2014mwa} which is similar to the procedure for rational CFTs, albeit with slight modification.
First we define the following matrix elements $F_{\alpha_s \alpha_t}$ (similar to the F matrix in Eq.(2.10) in \cite{Moore:1988ss}) by
\ba
F(\alpha_s|1-z)&=&\int_{\alpha_t}F_{\alpha_s\alpha_t}\cdot F(\alpha_t|z),\label{Ftr}
\ea
{where $F(\alpha|z)$ is the conformal block for the four-point function $\lla V_{\bar \alpha}(z_1,\bar{z}_1)
\ddd  V_{\alpha}(z_4,\bar{z}_4)\llb$. One should note that the fusion matrix is of infinite dimension which is different from that in rational CFTs.

%The prime stands for contour integral over continuous spectrum $n$ with some discrete terms $D_m$ which defined previously. As in presented in section, $D_m$ contains two contributions. The first one is just residues of contour integral which can cancel the contribution from the double poles of $C(\bar\alpha, \alpha,\alpha_s)C(\bar\alpha, \alpha,\bar\alpha_s)$ and the second one is that just factors with dropping out the divergent factor $C(\bar\alpha, \alpha,\alpha_s)C(\bar\alpha, \alpha,\bar\alpha_s)$. All these discrete terms are liking discrete spectrum as in rational CFTs. Therefore, we can do the analogous calculation to prove the variation of $n$th REE is still log of quantum dimension of local operator.

In the late time limit, the dominant contribution from the intermediate channel in Eq.(\ref{Ftr}) is denoted by $F_{\alpha}$ and it is defined as follows:
\bea
&&F_{\alpha}\sim F^{L, SL}_{\alpha_s=Q/2,\alpha_t=Q/2}\big[ {}_{\xa}^{\bar\xa} {}_{\bar\xa}^{\xa} \big] \nonumber\\&\times&
\left\{
\begin{array}{cc}
 (1-z)^{\Delta_{Q/2 }} \bar{z}^{\Delta_{Q/2 }} \ln^{-{3/2}}\left({1\over { (1-z)}}\right) \ln^{-{3/2}}\left({1\over { \bar z}}\right) & \alpha \in \{\alpha|\alpha=Q/2+ i p,  p\in  \mathbb{R}\}  \\&\bigcup \{\alpha|Q/2>\text{Re}(\alpha)> {Q/4}\}\bigcup \{\alpha|{Q/4>\text{Re}(\alpha)> {0}}\},\\
 (1-z)^{\Delta_{Q/2 }}  \bar{z}^{\Delta_{Q/2 }}\ln^{-{1\over 2}}\left({1\over { (1-z)}}\right)\ln^{-{3\over 2}}\left({1\over { \bar z}}\right)  & \alpha\in\{\alpha|\text{Re}(\alpha)=Q/4, \text{Im}(\alpha)\neq 0\}.\nonumber\\
\end{array}\right.\quad\quad\\&&
 \label{FO}
\eea
where $\sim$ denotes  neglect of the normalisation factors of two-point functions and the factors associated with structure constants $C(\alpha_1,\alpha_2,\alpha_3)$ and $C_{NS}(\alpha_1,\alpha_2,\alpha_3)$ in LFT and SLFT respectively\footnote{Refer to Eq.(\ref{LFT-late-1})Eq.(\ref{LFT-late-4}) in LFT and  Eq.(\ref{SLFT-late-1L})Eq.(\ref{SLFT-late-4L}) in SLFT. All these factors are divergent in the late time limit, namely $(z,\bar z)\to (1,0)$.}.

The $n$-th R\'enyi entanglement entropy can be obtained from the formula (\ref{replica}).
We find
\ba
&& \lla V_{\bar \alpha}(w_1,\bar{w}_1)V_{\alpha}(w_2,\bar{w}_2)\ddd V_{\alpha}(w_{2n},\bar{w}_{2n})\llb_{\Sigma_n} \no
&=& n^{-4n\Delta}\cdot (rs)^{-2(n-1)\Delta}\cdot \lla V_{\bar\alpha}(z_1,\bar{z}_1)V_{\alpha}(z_2,\bar{z}_2)\ddd V_{\alpha}(z_{2n},\bar{z}_{2n})\llb_{\Sigma_1},
\ea
where we define
\be
|z_{2k+1}|^n=r,\ \ \ |z_{2k+2}|^n=s.
\ee

We normalise the two point function\footnote{For the sake of simplicity, we have omitted the normalisation factor associated with $\delta(0)$ which does not affect the final conclusions.}
\ba
\lla V_{\bar \alpha}(w_1,\bar{w}_1)V_{\alpha}(w_2,\bar{w}_2)\llb_{\Sigma_1}
=\f{1}{|w_{12}|^{4\Delta}}=\f{1}{(2\ep)^{4\Delta}}. \label{ntw}
\ea

Then we get
\ba
&& \f{\lla V_{\bar \alpha}(w_1,\bar{w}_1)V_{\alpha}(w_2,\bar{w}_2)\ddd V_{\alpha}(w_{2n},\bar{w}_{2n})\llb_{\Sigma_n}}{\left(\lla V_{\bar \alpha}(w_1,\bar{w}_1)V_{\alpha}(w_2,\bar{w}_2)\llb_{\Sigma_1}\right)^n} \no
&=& \left(\f{2\ep}{n}\right)^{4\Delta n}\cdot (rs)^{-2(n-1)\Delta}\cdot \lla V_{\bar \alpha}(z_1,\bar{z}_1)V_{\alpha}(z_2,\bar{z}_2)\ddd V_{\alpha}(z_{2n},\bar{z}_{2n})\llb_{\Sigma_1}\no
&\to&  \left(\f{2\ep}{nt^{\f{n-1}{n}}}\right)^{4\Delta n}\cdot \lla V_{\alpha}(z_1,\bar{z}_1)V_{\alpha}(z_2,\bar{z}_2)\ddd V_{\alpha}(z_{2n},\bar{z}_{2n})\llb_{\Sigma_1},
\ea
where we take the late time limit in the final expression.

The $2n$ points $z_1, z_2,\ddd,z_n$ in the $z$ coordinate are given by
\ba
&& z_{2k+1}=e^{2\pi i\f{k}{n}}(i\ep+t-l)^{\f{1}{n}}=e^{2\pi i\f{k+1/2}{n}}(l-t-i\ep)^{\f{1}{n}} \no
&& z_{2k+2}=e^{2\pi i\f{k}{n}}(-i\ep+t-l)^{\f{1}{n}}=e^{2\pi i\f{k+1/2}{n}}(l-t+i\ep)^{\f{1}{n}},
\no
&& \bar{z}_{2k+1}=e^{-2\pi i\f{k}{n}}(-i\ep-t-l)^{\f{1}{n}}=e^{-2\pi i\f{k+1/2}{n}}(l+t+i\ep)^{\f{1}{n}} \no
&& \bar{z}_{2k+2}=e^{-2\pi i\f{k}{n}}(i\ep-t-l)^{\f{1}{n}}=e^{-2\pi i\f{k+1/2}{n}}(l+t-i\ep)^{\f{1}{n}}.
\ea

In the early time limit $t\ll l$ we find
\be
z_{2k+1}\to z_{2k+2},\ \ \ \bar{z}_{2k+1}\to \bar{z}_{2k+2}, \label{lmit}
\ee
for all $k$.
On the other hand, if we take the late time limit $t\gg l$, we find the asymmetric limit:
\be
z_{2k+1}\to z_{2k+4},\ \ \ \bar{z}_{2k+1}\to \bar{z}_{2k+2}, \label{mit}
\ee
for all $k$.

If we can regard the $2n$ point functions as $n$ products of two-point functions
(\ref{ntw}) in the late time limit, we have
\ba
&& \lla V_{\bar \alpha}(z_1,\bar{z}_1)\ddd V_{\alpha}(z_n,\bar{z}_n)\llb_{\Sigma_1} \no
& \to &
\lla V_{\bar \alpha}{(z_1,\bar{z}_1)}V_{\alpha}{(z_2,\bar{z}_2)}\llb_{\Sigma_1}\otimes \lla V_{\bar \alpha}{(z_3,\bar{z}_3)}V_{\alpha}{(z_4,\bar{z}_4)}\llb_{\Sigma_1}\otimes \ddd
\otimes\lla V_{\bar \alpha}{(z_{2n-1},\bar{z}_{2n-1})}V_{\alpha}{(z_{2n},\bar{z}_{2n})}\llb_{\Sigma_1},\quad\quad\nonumber\\
\ea
which respects the late limit in the anti-holomorphic sectors and $\otimes$ denotes that we only consider the dominant contribution with divergent factor. In the holomorphic sector, we would like to take the late time limit. To this end, we need to exchange some of $z_i$s with $z_j$s as
\be
(z_1,z_2)(z_3,z_4)\ddd (z_{2n-1},z_{2n})\to (z_1,z_4)(z_3,z_6)\ddd (z_{2n-1},z_{2}).
\ee
 This transformation is realized by applying the F-transformation $n-1$ times.

We can estimate the difference in the late time limit as follows:
\ba
&& z_{2k+1}-z_{2k+4}\simeq -e^{2\pi i\f{k+1/2}{n}}\cdot \f{2i\ep t^{(1/n-1)}}{n}. \no
&& \bar{z}_{2k+1}- \bar{z}_{2k+2}\simeq e^{-2\pi i\f{k+1/2}{n}}\cdot \f{2i\ep t^{(1/n-1)}}{n}.
\ea
Their absolute values are all the same and are given by
\be
\delta\equiv \f{2\ep}{nt^{\f{n-1}{n}}}.
\ee

Thus we can estimate the $2n$ point function in the late time limit as follows:
\ba
&& \lla V_{\bar \alpha}(z_1,\bar{z}_1)V_{\alpha}(z_2,\bar{z}_2)\ddd V_{\alpha}(z_{2n},\bar{z}_{2n})\llb_{\Sigma_1} \no
& \simeq &  (F_{\alpha})^{n-1}\cdot \left[\lim_{|z_{2k+1}-z_{2k+2}|=\delta\to 0}
\ \ \lla V_{\bar \alpha}(z_1,\bar{z}_1)V_{\alpha}(z_2,\bar{z}_2)\ddd V_{\alpha}(z_{2n},\bar{z}_{2n})\llb_{\Sigma_1} \right]
\no
& \simeq& (F_{\alpha})^{n-1}\cdot  \delta ^{-4\Delta n}.
\ea

Then
\be
\f{\lla V_{\bar \alpha}(w_1,\bar{w}_1)V_{\alpha}(w_2,\bar{w}_2)\ddd V_{\alpha}(w_{2n},\bar{w}_{2n})\llb_{\Sigma_n}}{\left(\lla V_{\bar \alpha}(w_1,\bar{w}_1)V_{\alpha}(w_2,\bar{w}_2)\llb_{\Sigma_1}\right)^n}=(F_{\alpha})^{n-1}.
\ee
The late time limit of the difference of the $n$th REE $\Delta {S}^{(n)}_{A}\big[ V_{\alpha}|0\rangle, V_{\alpha_r}|0\rangle\big]$ is
\bea
&&\Delta {S}^{(n)}_{A}\big[ V_{\alpha}|0\rangle, V_{\alpha_r}|0\rangle\big](t\to \infty)\nonumber\\&=&\left
 \{
\begin{array}{cc}
 -\log\Big({{ f_{\alpha}''(p){F^L_{Q/2,Q/2}}\big[ {}_{\xa}^{\bar\xa} {}_{\bar\xa}^{\xa} \big]}\over {f_{\alpha_r}''(p){F^L_{Q/2,Q/2}}\big[ {}_{\xa_r}^{\bar{\xa_r}} {}_{\bar{\xa_r}}^{\xa_r} \big]}}\Big) \Big|_{p\to 0}& \alpha \in \{\alpha|\alpha=Q/2+ i p|  p\in  \mathbb{R}\}\bigcup \{Q/2>\text{Re}(\alpha)> {Q/4}\}  \\&\bigcup \{{Q/4>\text{Re}(\alpha)> {0}}\},\\
  -\log\Big({{ f_{\alpha}(p){F^L_{Q/2,Q/2}}\big[ {}_{\xa}^{\bar\xa} {}_{\bar\xa}^{\xa} \big]}\over {f_{\alpha_r}(p){F^L_{Q/2,Q/2}}\big[ {}_{\xa_r}^{\bar{\xa_r}} {}_{\bar{\xa_r}}^{\xa_r} \big]}}\Big) \Big|_{p\to 0} & \alpha\in\{\alpha|\text{Re}(\alpha)=Q/4, \text{Im}(\alpha)\neq 0\}. \\
\end{array}\right.
\eea

Then the variation of the difference $\Delta {S}^{(n)}_{A}\big[ V_{\alpha}|0\rangle, V_{\alpha_r}|0\rangle\big]$ between early time and late time limit is
\bea\label{ratioofF}
&&\Delta {S}^{(n)}_{A}\big[ V_{\alpha}|0\rangle, V_{\alpha_r}|0\rangle\big](t\to \infty)-\Delta {S}^{(n)}_{A}\big[ V_{\alpha}|0\rangle, V_{\alpha_r}|0\rangle\big](t\to 0)\nonumber\\&=&\left
 \{
\begin{array}{cc}
 -\log\Big({{ {F^L_{Q/2,Q/2}}\big[ {}_{\xa}^{\bar\xa} {}_{\bar\xa}^{\xa} \big]}\over {{F^L_{Q/2,Q/2}}\big[ {}_{\xa_r}^{\bar{\xa_r}} {}_{\bar{\xa_r}}^{\xa_r} \big]}}\Big) \Big|_{p\to 0}& \alpha \in \{\alpha|\alpha=Q/2+ i p,  p\in  \mathbb{R}\}\bigcup \{Q/2>\text{Re}(\alpha)> {Q/4}\}  \\&\bigcup \{{Q/4>\text{Re}(\alpha)> {0}}\},\\
  -\log\Big({{ {F^L_{Q/2,Q/2}}\big[ {}_{\xa}^{\bar\xa} {}_{\bar\xa}^{\xa} \big]}\over {{F^L_{Q/2,Q/2}}\big[ {}_{\xa_r}^{\bar{\xa_r}} {}_{\bar{\xa_r}}^{\xa_r} \big]}}\Big) \Big|_{p\to 0} & \alpha\in\{\alpha|\text{Re}(\alpha)=Q/4, \text{Im}(\alpha)\neq 0\}. \\
\end{array}\right.
\eea
Here we have restored the structure constants in $F_{\alpha}$ defined by Eq.(\ref{ratioofF}). {This is the main difference between LFT and rational CFTs. {In Appendix \ref{fusionLFT} and \ref{fusionSLFT}, unlike that in rational CFTs, one can show that Eq.(\ref{FO}) cannot be identified as a quantum dimension \cite{He:2014mwa} which is associated with modular invariance \cite{Guo:2018lqq}. Therefore, the Eq.(\ref{ratioofF}) is no longer associated with topological quantity.}

{To close this section, we would like to comment on how to extend the calculation of $\Delta {S}^{(n)}_{A}\big[ V_{\alpha}|0\rangle, V_{\alpha_r}|0\rangle\big](t\to \infty)-\Delta {S}^{(n)}_{A}\big[ V_{\alpha}|0\rangle, V_{\alpha_r}|0\rangle\big](t\to 0)$ in generic CFTs. In terms of the main result Eq.(\ref{ratioofF}), $\Delta {S}^{(n)}_{A}\big[ V_{\alpha}|0\rangle, V_{\alpha_r}|0\rangle\big](t\to \infty)-\Delta {S}^{(n)}_{A}\big[ V_{\alpha}|0\rangle, V_{\alpha_r}|0\rangle\big](t\to 0)$ depends on the ratio of the fusion matrix elements associated with target states and reference states.  Technically speaking, these fusion matrix elements come from the intermediate dominant channel in late time limit.  Then the $\Delta {S}^{(n)}_{A}\big[ V_{\alpha}|0\rangle, V_{\alpha_r}|0\rangle\big](t\to \infty)-\Delta {S}^{(n)}_{A}\big[ V_{\alpha}|0\rangle, V_{\alpha_r}|0\rangle\big](t\to 0)$ can be obtained directly in generic CFTs. In this sense, to obtain the memory effect of REE in generic CFTs, we only care about the fusion matrix element associated with intermediate dominant holomorphic channel in late time limit.}
\section{The $n$th REE for generic descendent states}
{In the previous sections, we computed the $n$th REE of two classes of local excited states in LFT and SLFT respectively.
In this section, we would like to extend the analysis to states generated by generic descendent operators in LFT following \cite{Chen:2015usa}.}
In the late time limit, the authors of \cite{Chen:2015usa} show that the difference of the $n$-th REE only depends on the most singular term in the two-point function and the $2n$-point function. One can define the following generic descendent operator:
\ba \text{ } \tilde{ V}(w,\bar{w})=L^{(-)}V_{\alpha}(w,\bar{w}),\text{ } \ea
where $L^{(-)}$ is a  complicated product form of holomorphic generators, $V_{\alpha}(w,\bar{w})$ is a primary operator of conformal dimension $h$ and $\tilde{V}(w,\bar{w})$ is a quasi-primary operator. The operator $L^{(-)}$ has a fixed conformal dimension $m$
\ba \text{ }[L_0,L^{(-)}]=mL^{(-)}. \text{ }\ea

The conformal transformation for the descendant operators can be derived from the energy momentum tensor and the conformal transformation from the $z$ plane to $w$ plane is
\bea L_{-m}\mid_{w_1} &=&L_{-m}\mid_{z_1}\left(\frac{\partial z_1}{\partial w_1}\right)^n+...
\eea
where the ellipsis denotes the terms of lower conformal dimensions leading to less divergent terms in the limit $\epsilon\rightarrow0$.
By the conformal transformation, the four-point function is transformed as follows
\bea\label{tra}
&&{\langle \tilde{ V}^+(w_1,\bar{w}_1)\tilde{ V}(w^{'}_1,\bar{w}_1^{'})\tilde{ V}^+(w_2,\bar{w}_2)\tilde{ V}(w_2^{'},\bar{w}_2^{'})\rangle} \nonumber \\
&=&\left(\frac{\partial w_1}{\partial z_1}\right)^{h+m}\left(\frac{\partial \bar{w}_1}{\partial \bar{z}_1}\right)^{h+\bar{m}}
\left(\frac{\partial w_2}{\partial z_2}\right)^{h+m}\left(\frac{\partial \bar{w}_2}{\partial \bar{z}_2}\right)^{h+\bar{m}}
\left(\frac{\partial w_1^{'}}{\partial z_1^{'}}\right)^{h+m}\left(\frac{\partial \bar{w}_1^{'}}{\partial \bar{z}_1^{'}}\right)^{h+\bar{m}}
\left(\frac{\partial w_2^{'}}{\partial z_2^{'}}\right)^{h+m}\left(\frac{\partial \bar{w}_2^{'}}{\partial \bar{z}_2^{'}}\right)^{h+\bar{m}}
\nonumber \\
&~&\langle \tilde{ V}^+(z_1,\bar{z}_1)\tilde{ V}(z^{'}_1,\bar{z}_1^{'})\tilde{ V}^+(z_2,\bar{z}_2)\tilde{ V}(z_2^{'},\bar{z}_2^{'})\rangle
+\mbox{less divergent terms}.\text{ }\text{ }
\eea
The coefficient for the leading term is the same as that for the primary operator.
The terms with lower conformal dimensions do not change the final result.

The  two-point function for $V$ can be expressed as follows
\ba\label{twopoint}\text{ }\text{ } \langle (L^{(-)}V_{\alpha})^+(z),L^{(-)}V_{\alpha}(z^{'}) \rangle =\frac{(-1)^m\langle h \mid L^{(-)\dag}L^{(-)} \mid h \rangle}{(z-z^{'})^{2(h+m)}}.\text{ }\text{ } \ea

In the late time limit, $(z_1,z_2^{'})$ $(z_2,z_1^{'})$ $(\bar{z}_1,\bar{z}_1^{'})$ $(\bar{z}_2,\bar{z}_2^{'})$ approach each other.
The four-point correlation function of $V$ can be transformed to
\bea \text{}&&\text{}{\langle \tilde{ V}^+(z_1,\bar{z}_1)\tilde{ V}(z_1^{'},\bar{z}_1^{'})\tilde{ V}^+(z_2,\bar{z}_2)\tilde{ V}(z_2^{'},\bar{z}_2^{'}) \rangle }\text{}\text{}\nonumber \\
&=&{\cal{D}}
\langle V_{\bar \alpha}(z_1,\bar{z}_1)V_{\alpha}(z_1^{'},\bar{z}_1^{'})V_{\bar \alpha}(z_2,\bar{z}_2)V_{\alpha}(z_2^{'},\bar{z}_2^{'}) \rangle \text{ }\nonumber \\
&=&{\cal{D}}
\int'_m c_{m} \langle V_{\bar \alpha}(z_1)V_{\alpha}(z_1^{'})\mid_m V_{\bar \alpha}(z_2)V_{\alpha}(z_2^{'}) \rangle
\langle V_{\bar \alpha}(\bar{z}_1)V_{\alpha}(\bar{z}_1^{'})\mid_m V_{\bar \alpha}(\bar{z}_2) V_{\alpha}(\bar{z}_2^{'}) \rangle \text{}\nonumber \\
&=& {\cal{D}}
\int'_{m,n} c_{m,n} \langle V_{\bar \alpha}(z_1)V_{\alpha}(z_2^{'})\mid_m V_{\bar \alpha}(z_2)V_{\alpha}(z_1^{'}) \rangle
\langle V_{\bar \alpha}(\bar{z}_1)V_{\alpha}(\bar{z}_1^{'})\mid_n V_{\bar \alpha}(\bar{z}_2) V_{\alpha}(\bar{z}_2^{'}) \rangle \text{} \nonumber \\
&=&\text{}\int'_{m,n} c_{m,n} \langle L^{(-)}V_{\alpha}^+(z_1)L^{(-)}V_{\alpha}(z_2^{'})\mid_m L^{(-)}V_{\alpha}^+(z_2)L^{(-)}V_{\alpha}(z_1^{'}) \rangle
\langle V_{\bar \alpha}(\bar{z}_1)V_{\alpha}(\bar{z}_1^{'})\mid_n V_{\bar \alpha}(\bar{z}_2)V_{\alpha}(\bar{z}_2^{'}) \rangle.\text{} \nonumber \\\label{difference}
\eea
Here $\langle V_{\bar \alpha}(z_1)V_{\alpha}(z_1^{'})\mid_m V_{\bar \alpha}(z_2)V_{\alpha}(z_2^{'})\rangle$ denotes the conformal block expansion with the Virasoro module $[m]$ which exhibits continuous spectrum in Liouville field theory and $[m]$ contains the dominant contribution satisfying the fusion rule\footnote{Due to the fusion rule in LFT or SLFT, $[m]$ is not vacuum module which does not belong to the spectrum of LFT or SLFT.}. The correlation function of four descendant operators has been transformed into the differential operator ${\cal{D}}$ acting on the correlation function of corresponding primaries in the first equality. In the second equality, the partition function is expanded by the conformal blocks and here $c_m$ denotes the OPE coefficients. In the third equality, the holomorphic part can be expressed by $t$-channel like \cite{He:2014mwa} in the late time limit. In the fourth equality, we pull the differential operator back into the Virasoro operators acting on the primaries in the correlation function. In LFT and SLFT, we have shown that the dominant contribution to REE in late time limit comes from the intermediate channel with $m=Q/2, n=Q/2$ in primed contour integration over $m,n$ in Eq.(\ref{difference}).

In Sections 2 and 3, we have already found that the most divergent term comes from the intermediate operator with {the minimal conformal dimension of the operator in the intermediate fusion channel} and $c_{\text{Min}(\alpha_s),\text{Min}(\alpha_t)}={\lb F^L_{\text{Min}(\alpha_s),\text{Min}(\alpha_t)}\big[ {}_{\xa}^{\bar\xa} {}_{\bar\xa}^{\xa} \big]\rb}^{-1}={1\over F_{\alpha}^{L,SL}}$\footnote{Here the $F_{\alpha}^{L,SL}$ contains the divergent piece shown in eq.(\ref{FO}) and $\text{Min}(\alpha_s), \text{Min}(\alpha_t)$ are the minimal conformal dimension of intermediate operator involving in the fusion procees.} which is similar to that in rational CFTs \cite{He:2014mwa}. We use upper index $L, SL$ to distinguish the quantities in LFT and SLFT. The most divergent term is as follows,
\bea &&{\langle \tilde{ V}^+(z_1,\bar{z}_1)\tilde{ V}(z_1^{'},\bar{z}_1^{'})\tilde{ V}^+(z_2,\bar{z}_2)\tilde{ V}(z_2^{'},\bar{z}_2^{'}) \rangle} \nonumber \\
&=& \frac{1}{F^{L,SL}_{\text{Min}(\alpha_s),\text{Min}(\alpha_t)}\big[ {}_{\xa}^{\bar\xa} {}_{\bar\xa}^{\xa} \big]}
\langle L^{(-)}V_{\alpha}^+(z_1)L^{(-)}V_{\alpha}(z_2^{'})\rangle \otimes\langle L^{(-)}V_{\alpha}^+(z_2)L^{(-)}V_{\alpha}(z_1^{'}) \rangle\otimes
\langle V_{\bar \alpha}(\bar{z}_1)V_{\alpha}(\bar{z}_1^{'}) \rangle\nonumber\\&~&\otimes \langle V_{\bar \alpha}(\bar{z}_2) V_{\alpha}(\bar{z}_2^{'}) \rangle
+\text{less divergent terms.}
\eea
So the four-point function in $w$-coordinate keeping the most divergent term is
\bea \text{}&&{\langle \tilde{ V}^+(w_1,\bar{w}_1)\tilde{ V}(w_1^{'},\bar{w}_1^{'})\tilde{ V}^+(w_2,\bar{w}_2)\tilde{ V}(w_2^{'},\bar{w}_2^{'}) \rangle }\text{ }\nonumber \\
&=& \frac{1}{F^{L,SL}_{\text{Min}(\alpha_s),\text{Min}(\alpha_t)}\big[ {}_{\xa}^{\bar\xa} {}_{\bar\xa}^{\xa} \big]}
\langle L^{(-)}V_{\alpha}^+(w_1)L^{(-)}V_{\alpha}(w_2^{'})\rangle \otimes\langle L^{(-)}V_{\alpha}^+(w_2)L^{(-)}V_{\alpha}(w_1^{'}) \rangle\otimes
\langle V_{\bar \alpha}(\bar{w}_1)V_{\alpha}(\bar{w}_1^{'}) \rangle\nonumber\\&~&\otimes \langle V_{\bar \alpha}(\bar{w}_2) V_{\alpha}(\bar{w}_2^{'})  \rangle +\text{less divergent terms.}\text{ }
\eea
Therefore, for a quasi-primary operator we still have ${S}^{(2)}[\tilde{ V}|0\rangle](t\to \infty)=-\log F^{L,SL}_{\text{Min}(\alpha_s),\text{Min}(\alpha_t)}\big[ {}_{\xa}^{\bar\xa} {}_{\bar\xa}^{\xa} \big] $\footnote{Precisely, here we have neglected the normalisation factors and associated DOZZ factors.} which is defined by Eq.(\ref{FO}) in LFT and SLFT.

Finally, we present the main results with the more generic descendent operators such as
\ba \text{}\tilde{V}=\sum_{m,j,r,k}d_{m,j,r,k}\partial^{m}L^{(-,j)}\text{ }\bar{\partial}^{r}\bar{L}^{(-,k)} V_{\alpha}(z,\bar{z})\label{Genericoperator} \text{} \ea
Here $j$ and $k$ denote the quasi-primary operators and $V_{\alpha}$ is primary operator. Where $L^{(-,j)}\bar{L}^{(-,k)}O_a(w,\bar{w})$ is a quasi-primary operator and $L^{(-,j)}$ ($\bar{L}^{(-,k)}$) is a combination of holomorphic (or anti-holomorphic) Virasoro generators with fixed conformal dimension $[L_0,L^{(-,j)}]=p_jL^{(-,j)}$.

{We do not repeat the calculation in detail here as this has been analyzed in \cite{Chen:2015usa}. The differences have been already presented in Eq.(\ref{difference}).
Here we provide the final result as follows
\ba\label{genericresults}\text{ }\text{ } {S}_{n}\big[ \tilde{V}|0\rangle\big](t\to \infty)&=&\log F^L_{\text{Min}(\alpha_s),\text{Min}(\alpha_t)}\big[ {}_{\xa}^{\bar\xa} {}_{\bar\xa}^{\xa} \big]-\frac{1}{n-1}\log \Tr \rho_0^{n}\nonumber\\&=& S_n^{\mbox{\tiny primary}}-\frac{1}{n-1}\log \Tr \rho_0^{n},\text{ }\text{ } \ea}
with the normalised density matrix
$\rho_0=\frac{\rho}{\Tr \rho}, $ where
\be
\text{ } {S}_n^{\mbox{\tiny primary}}\big[ {V_{\alpha}}|0\rangle\big](t\to \infty)=-\log F^L_{\text{Min}(\alpha_s),\text{Min}(\alpha_t)}\big[ {}_{\xa}^{\bar\xa} {}_{\bar\xa}^{\xa} \big]
\ee
is the REE of the local excited state $V_{\alpha}|0\rangle$. The density matrix is defined as follows:
\ba \text{ }\rho=BMB^{\dag}M^{\dag}, \text{ }\ea
and these matrices are associated with coefficients in Eq.(\ref{Genericoperator})
\ba B_{\{m,j\},\{r,k\}}&=&d^{*}_{m,j,r,k},\nonumber\\  M_{\{m,j\},\{r,k\}}&=&\langle h\mid L^{(-,j)\dag}L^{(-,j)} \mid h\rangle \delta_{j,k}i^{r-m}, \ea

As we see from Eq.(\ref{genericresults}), ${S}_{n}\big[ V_{\alpha}|0\rangle\big](t\to \infty)$ has a similar structure in LFT and SLFT with that in rational CFTs. ${S}_{n}\big[ V_{\alpha}|0\rangle\big](t\to \infty)$ makes two main contributions: the first contains the universal part depending on the fusion matrix element of the corresponding primary operator and it is divergent due to the divergent component Eq.(\ref{FO}) in the first term. The other comes from the normalisation scheme of local descendent operator. Finally, we can see that the variation of the difference $\Delta S_{EE}^{(n)}\big[ V_{\alpha}|0\rangle, V_{\alpha_r}|0\rangle\big]$ between the early time and the late time is
\bea &&\Delta S_{EE}^{(n)}\big[ V_{\alpha}|0\rangle, V_{\alpha_r}|0\rangle\big](t\to \infty)-\Delta S_{EE}^{(n)}\big[ V_{\alpha}|0\rangle, V_{\alpha_r}|0\rangle\big](t\to 0)\nonumber\\&=&
\left\{
\begin{array}{cc}
 -\log\Big({{ {F^L_{Q/2,Q/2}}\big[ {}_{\xa}^{\bar\xa} {}_{\bar\xa}^{\xa} \big]}\over {{F^L_{Q/2,Q/2}}\big[ {}_{\xa_r}^{\bar{\xa_r}} {}_{\bar{\xa_r}}^{\xa_r} \big]}}\Big) \Big|_{p\to 0}-\frac{1}{n-1}\log {tr \rho_0^{n}\over tr \tilde{\rho}_0^{n}} & \alpha \in \{\alpha|\alpha=Q/2+ i p, p\in  \mathbb{R}\}\bigcup \{Q/2>\text{Re}(\alpha)> {Q/4}\}  \\&\bigcup \{{Q/4>\text{Re}(\alpha)> {0}}\},\\
  -\log\Big({{ {F^L_{Q/2,Q/2}}\big[ {}_{\xa}^{\bar\xa} {}_{\bar\xa}^{\xa} \big]}\over {{F^L_{Q/2,Q/2}}\big[ {}_{\xa_r}^{\bar{\xa_r}} {}_{\bar{\xa_r}}^{\xa_r} \big]}}\Big) \Big|_{p\to 0}-\frac{1}{n-1}\log {tr \rho_0^{n}\over tr \tilde{\rho}_0^{n}}  & \alpha\in\{\alpha|\text{Re}(\alpha)=Q/4, \text{Im}(\alpha)\neq 0\}. \nonumber\\
\end{array}\right.\\ \label{differenceoftwo}\eea
where we choose $\tilde{\rho}_0$ defined by the reference state (\ref{Genericoperator}) associated with primary operator $V_{\alpha_r}$.

\section{Discussions and conclusions}
{In this paper, the time evolution of the difference $\Delta S_{EE}^{(n)}\big[ V_{\alpha}|0\rangle, V_{\alpha_r}|0\rangle\big](t)$ of REE  between locally excited states $V_{\alpha}|0\rangle$ and the reference states $V_{\alpha_r}|0\rangle$ has been studied in $1+1$ dimensional irrational CFT, especially in Liouville field theory and super Liouville field theory.}  In rational CFTs, there are {a finite number of} primary operators and one needs to undertake the finite summation to extract the difference of REE shown in \cite{He:2014mwa}. Furthermore the reference states $V_{\alpha_r}|0\rangle$ can be chosen as a vacuum state $\mathbf{1}|0\rangle$. In irrational CFTs, there are infinite dimensional and continuous spectra which are also highly degenerate. One might doubt that the differences of REE in irrational CFTs have very different structures compared with that \cite{He:2014mwa} in rational CFTs.

To answer this question, {we calculate the second REE in a compact $c=1$ boson at generic radius which is a irrational CFT as a preliminary exercise}. Because the Hilbert spaces of LFT and SLFT are very special in irrational CFTs, they are especially chosen as playgrounds to calculate the difference of REE $\Delta S_{EE}^{(n)}\big[ V_{\alpha}|0\rangle, V_{\alpha_r}|0\rangle\big](t)$ between two excited states, e.g. $ V_{\alpha_r}|0\rangle$ and $V_{\alpha_r}|0\rangle$.} Furthermore, there is no well studied holographic dual of local primary operator in integrable rational CFTs \cite{Gaberdiel:2012uj} and one cannot make use of AdS/CFT to calculate REE in this case. Although the holographic dual of LFT or SLFT is not clear, it is interesting to extract in the large c universal properties thereof. One can study time evolution of REE $ S_{EE}^{(n)}\big[ V_{\alpha}|0\rangle\big]$ by applying the replica trick in LFT and SLFT and one might obtain some properties of large c CFTs.

{To understand these properties, the second REE $ S_{EE}^{(2)}\big[ V_{\alpha}|0\rangle\big]$ of local excited states is calculated in LFT and SLFT. For a state excited by a local primary operator $V_\alpha$, the REE is divergent both in the early and late time limits. The divergent behaviours of REE $ S_{EE}^{(2)}\big[ V_{\alpha}|0\rangle\big](t)$ in the early and late time  limits are different, which seems to contradict the quasi-particle picture proposed in rational CFTs \cite{Nozaki:2014hna}\cite{He:2014mwa}. That also means that $S_{EE}^{(2)}\big[ V_{\alpha}|0\rangle\big](t\to \infty)-S_{EE}^{(2)}\big[ V_{\alpha}|0\rangle\big](t\to 0)$ is divergent \footnote{Equivalently, $\Delta {S}^{(n)}_{A}\big[ V_{\alpha}|0\rangle, 1|0\rangle\big](t\to \infty)-\Delta {S}^{(n)}_{A}\big[ V_{\alpha}|0\rangle, 1|0\rangle\big](t\to 0)$ is divergent.}. The identity operator does not live in the Hilbert space of LFT and SLFT and no discrete terms contribute to the REE, the vacuum block does not make a contribution to REE. That is the main reason leading to the different divergent behavior of REE in the early and late time limits. To define finite quantities, e.g. $\Delta {S}^{(n)}_{A}\big[ V_{\alpha}|0\rangle, V_{\alpha_r}|0\rangle\big](t)$, one has to classify all locally excited states in LFT and SLFT. The zero point of the structure constant (DOZZ formula) presented in the second REE has been estimated to classify the primary operators in LFT and SLFT.} These primary operators have been divided into two classes in terms of real part of Liouville momentum $\alpha$, e.g. $\alpha \in\{\alpha|\alpha=Q/2+ i p,  p\in  \mathbb{R}\}\bigcup \{\alpha|Q/2>\text{Re}(\alpha)> {Q/4}\}\bigcup \{\alpha|{Q/4>\text{Re}(\alpha)> {0}}\}$ and $\alpha \in\{\alpha|\text{Re}(\alpha)=Q/4, \text{Im}(\alpha)\neq 0\}$. Due to the fact that the second REE of excited states is divergent, one has to choose an appropriate reference state $ V_{\alpha_r}|0\rangle$ which lives in the same class as the target states $V_{\alpha}|0\rangle$. The difference $\Delta {S}^{(2)}_{A}\big[ V_{\alpha}|0\rangle, V_{\alpha_r}|0\rangle\big](t)$ of the second REE between target states $V_{\alpha}|0\rangle$ and reference states $V_{\alpha_r}|0\rangle$ will be finite in both the early and late limits. One can study the time evolution behaviour of $\Delta {S}^{(2)}_{A}\big[ V_{\alpha}|0\rangle, V_{\alpha_r}|0\rangle\big](t)$. The difference of the REE $\Delta {S}^{(2)}_{A}\big[ V_{\alpha}|0\rangle, V_{\alpha_r}|0\rangle\big]$ between the early time limit and the late time limit always coincides with the log of the ratio ${F_{\alpha_s=Q/2,\alpha_t=Q/2}\big[ {}_{\xa}^{\bar\xa} {}_{\bar\xa}^{\xa} \big]\over F_{\alpha_s=Q/2,\alpha_t=Q/2}\big[ {}_{\xa_r}^{\bar\xa_r} {}_{\bar\xa_r}^{\xa_r} \big]}$ of fusion matrix element between two excited states e.g. $V_{\alpha}|0\rangle, V_{\alpha_r}|0\rangle$. The precise expression has been listed in the Eq.(\ref{differenceoftwo}).
Following \cite{He:2014mwa}, one can also directly extend this analysis to the generic $n$-th REE. The difference $\Delta {S}^{(n)}_{A}\big[ V_{\alpha}|0\rangle, V_{\alpha_r}|0\rangle\big](t)$ between the early and late time is independent of $n$ and it still contains log of the ratio of fusion matrix elements ${F_{\alpha_s=Q/2,\alpha_t=Q/2}\big[ {}_{\xa}^{\bar\xa} {}_{\bar\xa}^{\xa} \big]\over F_{\alpha_s=Q/2,\alpha_t=Q/2}\big[ {}_{\xa_r}^{\bar\xa_r} {}_{\bar\xa_r}^{\xa_r} \big]}$ between the target and reference state. Finally, REE of the generic descendent states following \cite{Chen:2015usa} has been investigated. Comparing this with the case of primary states, the difference $\Delta {S}^{(n)}_{A}\big[ \tilde{V}_{\alpha}|0\rangle, \tilde{V}_{\alpha_r}|0\rangle\big](t)$ in descendent states $\tilde{V}_{\alpha}|0\rangle, \tilde{V}_{\alpha_r}|0\rangle$ in LFT or SLFT now contains one more additional terms which is associated with the normalisation factor of the descendent operator.

{How to understand the different divergence behavior of $\Delta {S}^{(n)}_{A}\big[ \tilde{V}_{\alpha}|0\rangle, \tilde{V}_{\alpha_r}|0\rangle\big]$ between the two classes of local excited states? In terms of \cite{Seiberg:1990eb}, they define the states $V_\alpha |0\rangle$ as Liouville momentum $\alpha \in\{Q/2+ ip, p\in \mathbb{R}\}$ is normalisable and states $V_\alpha |0\rangle$ with $0<\alpha<Q/2$ are non-normalisable states. The normalisable states correspond to non-local operators which create macroscopic holes in the surface. In this paper, we have confirmed the difference between the normalisable states and non-normalisable states from the memory effect of the REE perspective.}

{Finally, one can apply these techniques to calculate the out of time ordered correlation function (OTOC) to check that whether the super-integrability of LFT is consistent with the chaotic proposal \cite{Roberts:2014ifa}. Recently, the authors of \cite{Mertens:2017mtv} proposed a correspondence between 1+1 d Liouville field theory and the one dimensional conformal quantum mechanics SYK model. One can compare the OTOC in Liouville field theory with the late time behaviour of two-point function of bi-local operators in SYK and then check the correspondence \cite{Mertens:2017mtv} between LFT and SYK. More recently, the authors of \cite{Caputa:2017urj}\cite{Czech:2017ryf}\cite{Caputa:2017yrh} proposed that Liouville field theory action as the optimisation of the complexity of static states in conformal field theory. Also the associated measurements of the complexity in generic field theory and holographic aspects thereof have also been proposed in \cite{Jefferson:2017sdb}\cite{Chapman:2017rqy} and \cite{Susskind:2014rva}\cite{Alishahiha:2015rta} \cite{Brown:2015lvg} \cite{Brown:2015bva}\cite{Abt:2017pmf} respectively. One can extend the study of the complexity of local excited states in conformal field theory to define the optimisation procedure. Hopefully, some progress in these directions can be reported in the near future.}
\section*{Acknowledgements}

We would like to thank Konstantin Aleshkin, Vladimir Belavin, X.~Cao, Bin Chen, Harald Dorn, M.~R.~Gaberdiel, Wu-Zhong Guo, George Jorjadze, Li Li, Zhu-Xi~Luo, Tokiro Numasawa, Hao-Yu~Sun, J.~Teschner, Kento Watanabe, Jie-qiang Wu for their discussions and suggestions during various stages of the project. We appreciate Harald Dorn, Xing Huang, Axel Kleinschmidt, Hermann Nicolai, Tadashi Takayanagi to comment the draft. We especially thanks to Xing Huang and Stefan Theisen for intensive discussions during the whole project. S.H. appreciate Axel Kleinschmidt, Hermann Nicolai, Stefan Theisen and Tadashi Takayanagi for their encouragements and supports. S.H. is supported from Max-Planck fellowship in Germany, the German-Israeli Foundation for Scientific Research and Development and the National Natural Science Foundation of China (No.11305235).

\section{Appendix}
\subsection{The Notations of LFT}\label{reviewLFT}
The full Liouville action (See review \cite{Nakayama:2004vk}) is
\ba
\label{liouvaction}
S_L=\frac{1}{4\pi} \int{d^2 \xi \sqrt{{g}}\left[\partial_a \phi \partial_b \phi
{g}^{ab}+Q {{R}}\phi+4 \pi \mu e^{2b\phi}\right]},
\ea
where $Q=b+{1\over b}$.
The conformal dimension of corresponding primary operator $e^{2 \alpha \phi}$ is
\ba
\label{dimension}
\Delta(e^{2 \alpha \phi}){}{}=\bar{\Delta}(e^{2 \alpha \phi})=\alpha (Q-\alpha),{}{}
\ea

The stress tensor is
\ba
\label{stresstensor}
T(z)=-(\partial \phi)^2+Q\partial^2 \phi,
\ea
and the central charge of the conformal algebra is
\ba
c_L=1+6Q^2=1+6(b+b^{-1})^2. \label{centralcharge}
\ea

The three point function of primary operator in LFT is
\ba
\label{3point}
\langle V_{\alpha_1}(z_1,\bar{z}_1)V_{\alpha_2}(z_2,\bar{z}_2){}{}
V_{\alpha_3}(z_3,\bar{z}_3)\rangle=
\frac{C(\alpha_1,\alpha_2,\alpha_3)}{|z_{12}|^{2(\Delta_1+\Delta_2-\Delta_3)}{}{}
|z_{13}|^{2(\Delta_1+\Delta_3-\Delta_2)}|z_{23}|^{2(\Delta_2+\Delta_3-\Delta_1)}},
\ea
with $z_{ij}=z_i-z_j$.  The function $C(\alpha_1,\alpha_2,\alpha_3)$
is called structure constants associated with dynamical data of any CFT. The
DOZZ formula is an analytic expression for $C$ in LFT by
\cite{Dorn:1994xn,Zamolodchikov:1995aa}.
The DOZZ formula gives the three-point function
\ba\label{dozz}
&&C(\alpha_1,\alpha_2,\alpha_3)=\lambda^{(Q-\sum_{i=1}^3\alpha_i)/b}\times\\ \nonumber
&&{\Upsilon'_b(0)\Upsilon_b(2\alpha_1)\Upsilon_b(2\alpha_2)\Upsilon_b(2\alpha_3)\over
\Upsilon_b(\alpha_1+\alpha_2+\alpha_3-Q)\Upsilon_b(\alpha_1+\alpha_2-\alpha_3)
\Upsilon_b(\alpha_2+\alpha_3-\alpha_1)\Upsilon_b(\alpha_3+\alpha_1-\alpha_2)}\, ,
\ea
where
\ba
\Upsilon_{b}(x)={1\over \Gamma_b(x) \Gamma_b(Q-x)}\, ,
\ea
and
\be
\lambda=\pi\mu\gamma\left(b^2\right)b^{2-2b^2}\, .
\ee
The $\Gamma_b(x)$ is given by eq.(\ref{Gamma}).
In this paper, we need four point function of primary operator which reads
\begin{align} \nonumber
\label{4point}
&\biggl\langle V_{\alpha_1}(z_1,\overline{z}_1)V_{\alpha_2}(z_2,\overline{z}_2)
V_{\alpha_3}(z_3,\overline{z}_3)V_{\alpha_4}(z_4,\overline{z}_4)\biggr\rangle \\
=&|z_{13}|^{2(\Delta_4-\Delta_1-\Delta_2-\Delta_3)}|z_{14}|^{2(\Delta_2+
\Delta_3-\Delta_1-\Delta_4)}
|z_{24}|^{-4 \Delta_2}|z_{34}|^{2(\Delta_1+\Delta_2-\Delta_3-\Delta_4)}
 {G}_{1234}(z,\bar{z}),
\end{align}
with the harmonic cross ratio $z$ defined as:
\be
\label{harmonicratio}
z=\frac{z_{12}z_{34}}{z_{13}z_{24}}.
\ee
and ${G}_{1234}(z,\bar{z})$
\begin{align}
\label{4point} \nonumber
{G}_{1234}(z,\bar{z})=\frac{1}{2}&\int_{-\infty}^{\infty}\frac{dp}{2\pi}
C(\alpha_1,\alpha_2,Q/2+ip)R_{L}(\alpha_s)C(\alpha_3,\alpha_4,Q/2-i p)\\
&\times {F}_{1234}(\Delta_i,\Delta_p,z){F}_{1234}(\Delta_i,\Delta_p,\bar{z}).
\end{align}
Here $\Delta_p=p^2+Q^2/4$, and the function ${F}_{1234}(z)$ and ${F}_{1234}(\bar{z})$ are the Virasoro conformal block. In this paper, we follow the notation for four point Green function given in \cite{McElgin:2007ak} with respect to normalization of two point function.
\subsection{The Notations of SLFT}\label{SLFT}
The $\mathcal{N}=1$ supersymmetric Liouville field theory
may be defined by the action (See review \cite{Nakayama:2004vk})
\ba{}{}
S_{SL}=\frac{1}{4\pi} \int{d^2 \xi \sqrt{{g}}}{1\over 2\pi}g_{ab}\partial_a\varphi\partial_b \varphi+{}{}
{1\over 2\pi}{}{}(\psi\bar{\partial}\psi+\bar{\psi}\partial\bar{\psi})+2i\mu b^2\bar{\psi}\psi e^{b\varphi}+{}{}
2\pi \mu^2 b^2 e^{2b\varphi}\, ,
\ea
where $\varphi$ is a bosonic and $\psi$ a fermionic field,
$\mu$ denotes a two-dimensional cosmological constant and $b$ is a Liouville coupling constant.

The theory has $\mathcal{N}=1$ superconformal symmetry. The energy-momentum tensor and the superconformal current are
\ba
T&=&-{1\over 2} {}{}{}(\partial\varphi\partial\varphi-Q\partial^2\varphi+\psi\partial\psi)\, ,\nonumber\\
G&=&i(\psi\partial\varphi-Q\partial\psi){}{}\nonumber.
\ea
and the superconformal algebra is
\ba
&& [L_m,L_n]=(m-n)L_{m+n}+{c\over 12}m(m^2-1)\delta_{m+n}\, ,\\
 &&[L_m, G_k]={m-2k\over 2}G_{m+k}\, ,\\
 &&\{G_k,G_l\}=2L_{l+k}+{c\over 3}\left(k^2-{1\over 4}\right)\delta_{k+l}\, .
\ea
The central charge in SLFT is given by
\be
{}{}c = \frac 3 2+ 3 Q^2{}{}{}.
\ee

The NS-NS primary fields $e^{\alpha \varphi(z,\bar{z})}$ in the $\mathcal{N}=1$ SLFT have conformal dimensions
\be
\Delta^{NS}_{\alpha}={1\over 2}\alpha(Q-\alpha){}{} .
\ee
As before, physical states have $\alpha={Q\over 2}+ip$ with $Q=b+{1\over b}$.
The R-R primary field is defined as
\ba
{}{}R^\xe_{\alpha}(z,\bar{z})=\sigma^\xe(z,\bar{z}) e^{\alpha \varphi(z,\bar{z})}\,{}{}{} ,
\ea
where $\sigma$ is the spin field and $\xe = \pm$ is the fermion parity. For simplicity we can take all $\xe = +$ and drop this index.

The dimension of the R-R operator is
\be
\Delta_{\alpha}^R={1\over 16}+{1\over 2}\alpha(Q-\alpha)\, .
\ee

To consider both NS and R sectors, we will need various functions defined differently for each sector. Here we use the notations $C_i, \Upsilon_i, \xG_i$, where $i = 1 \zt{ mod } 2$ for $C_\zt{NS}, \Upsilon_\zt{NS}, \xG_\zt{NS}$ and $i = 0 \zt{ mod } 2$ for R. One can refer to appendix \ref{doublesine} to find out the exact definition of these special functions.

The four point function for NS-NS operator is
\ba
G_4(z,\bar z)
=
\Big\langle
V_{\alpha_4}(\infty,\infty)
V_{\alpha_3}(1,1)
V_{\alpha_2}(z,\bar z)
V_{\alpha_1}(0,0)
\Big\rangle,
\ea
which can be written in the ``$s-$channel'' representation:
\ba\label{4pt-SLFT}
G_4(z,\bar z) & = &
\int_{\frac{Q}{2} + i{\mathbb R}^+}
\frac{d\alpha_s}{i}
\left[
C(\alpha_4,\alpha_3,\alpha_s)C(\bar\alpha_s,\alpha_2,\alpha_1)
\left|{ F}_{\alpha_s}^{\rm e}\left[^{\alpha_3\: \alpha_2}_{\alpha_4\: \alpha_1}\right](z)\right|^2
\right.
\\
&&
\left.
-\
\tilde C(\alpha_4,\alpha_3,\alpha_s)\tilde C(\bar\alpha_s,\alpha_2,\alpha_1)
\left|{ F}_{\alpha_s}^{\rm o}\left[^{\alpha_3\: \alpha_2}_{\alpha_4\: \alpha_1}\right](z)\right|^2
\right].
\ea
The e,o in ${ F}_{\alpha_s}^{\rm e}$ and ${ F}_{\alpha_s}^{\rm o}$ denote $\mathcal{N}=1$ Neveu-Schwarz blocks with even and odd fermion parity as in \cite{Hadasz:2006qb}\cite{Belavin:2007gz}.

Following \cite{Rubik, Marian} the structure constants have the following explicit form ($\alpha$
stands here for $\alpha_{1}+\alpha_{2}+\alpha_{3}$)
\ba
C({\alpha_{1},\alpha_{2},\alpha_{3}})  & =&\left(  \pi\mu\gamma\left(  \frac{Qb}2\right)
b^{1-b^{2}}\right)  ^{(Q-\alpha)/b}\frac{\Upsilon_{\text{NS}}^{\prime}
(0)\Upsilon_{\text{NS}}(2\alpha_{1})\Upsilon_{\text{NS}}(2\alpha_{2})\Upsilon
_{\text{NS}}(2\alpha_{3})}{\Upsilon_{\text{NS}}(\alpha-Q)\Upsilon_{\text{NS}}
(\alpha_{1+2-3})\Upsilon_{\text{NS}}(\alpha_{2+3-1})\Upsilon_{\text{NS}}(\alpha_{3+1-2}
)}\nonumber\\
\tilde C({\alpha_{1},\alpha_{2},\alpha_{3}})  & =&\left(  \pi\mu\gamma\left(  \frac{Qb}2\right)
b^{1-b^{2}}\right)  ^{(Q-\alpha)/b}\frac{2i\Upsilon_{\text{NS}}^{\prime}
(0)\Upsilon_{\text{NS}}(2\alpha_{1})\Upsilon_{\text{NS}}(2\alpha_{2})\Upsilon
_{\text{NS}}(2\alpha_{3})}{\Upsilon_{\text{R}}(\alpha-Q)\Upsilon_{\text{R}}
(\alpha_{1+2-3})\Upsilon_{\text{R}}(\alpha_{2+3-1})\Upsilon_{\text{R}}(\alpha_{3+1-2}
)}\nonumber\\\label{3ptfunctionSLFT}
\ea Where we define $\alpha_{i+j-k}=\alpha_i+\alpha_j-\alpha_k$ for short and
\be
\Upsilon_{1}(x)\equiv\Upsilon_{\rm NS}(x)=\Upsilon_{b}\left({x\over 2}\right)\Upsilon_{b}\left({x+Q\over 2}\right)={1\over \Gamma_{\rm NS}(x) \Gamma_{\rm NS}(Q-x)}\, ,
\ee

\be
\Gamma_{1}(x)\equiv\Gamma_{\rm NS}(x)=\Gamma_b\left({x\over 2}\right)\Gamma_b\left({x+Q\over 2}\right)\, .
\ee
%\be
%\lambda=\pi\mu\gamma\left({bQ\over 2}\right)b^{1-b^2}\, .
%\ee
Functions for R sector are defined differently. For example, we have
\be
\Gamma_{0}(x)\equiv\Gamma_{\rm R}(x)=\Gamma_b\left({x+b\over 2}\right)\Gamma_b\left({x+b^{-1}\over 2}\right)\, .
\ee

The four point function for R-R operators
\ba
G_4(z,\bar z)
= \Big\langle
R_{\alpha_4}(\infty,\infty)
R_{\alpha_3}(1,1)
R_{\alpha_2}(z,\bar z)
R_{\alpha_1}(0,0)
\Big\rangle,
\ea
can be also written in a similar form
\ba\label{4pt-SLFT}
G_4(z,\bar z) & = &
\int_{\frac{Q}{2} + i{\mathbb R}^+}
\frac{d\alpha_s}{i}
\left[
C^+_R(\alpha_4,\alpha_3|\alpha_s)C^+_R(\alpha_2,\alpha_1|\bar\alpha_s)
\left|{ F}_{\alpha_s}^{\rm e}\left[^{\alpha_3\: \alpha_2}_{\alpha_4\: \alpha_1}\right](z)\right|^2
\right.
\nn
&&
\left.
-\
\tilde C^+_R(\alpha_4,\alpha_3|\alpha_s)\tilde C^+_R(\alpha_2,\alpha_1|\bar\alpha_s)
\left|{ F}_{\alpha_s}^{\rm o}\left[^{\alpha_3\: \alpha_2}_{\alpha_4\: \alpha_1}\right](z)\right|^2
\right].
\ea
The corresponding structure constants become
\ba\label{dozzRR}
&&C^\xe_{R}(\alpha_1,\alpha_2|\alpha_3)=\left(  \pi\mu\gamma\left(  \frac{Qb}2\right)
b^{1-b^{2}}\right)  ^{(Q-\alpha)/b}\times \lsb {\Upsilon'_{NS}(0)\Upsilon_{R}(2\alpha_1)\Upsilon_{R}(2\alpha_2)\Upsilon_{NS}(2\alpha_3)\over
\Upsilon_{R}(\alpha-Q)\Upsilon_{R}(\alpha_{1+2-3})
\Upsilon_{NS}(\alpha_{2+3-1})\Upsilon_{NS}(\alpha_{3+1-2})}\rc \nn
&&+\lc {\xe \Upsilon'_{NS}(0)\Upsilon_{R}(2\alpha_1)\Upsilon_{R}(2\alpha_2)\Upsilon_{NS}(2\alpha_3)\over
\Upsilon_{NS}(\alpha-Q)\Upsilon_{NS}(\alpha_{1+2-3})
\Upsilon_R(\alpha_{2+3-1})\Upsilon_R(\alpha_{3+1-2})} \rsb \\
& & \tilde C^\xe_{R}(\alpha_1,\alpha_2|\alpha_3) = -\frac {i\xe} 2 [(p_1^2+p_2^2) C^\xe_{R}(\alpha_1,\alpha_2|\alpha_3) - 2 p_1 p_2 C^{-\xe}_{R}(\alpha_1,\alpha_2|\alpha_3)]\,.
\ea

\subsection{The function $\Gamma_b(x)$}
The function $\Gamma_b(x)$ is a close relative of the double Gamma function studied in \cite{Ba}. It
can be expressed by means of the integral representation
\ba\label{Gamma}
\log\Gamma_b(x)=\int_0^{\infty}\frac{dt}{t}
\biggl(\frac{e^{-xt}-e^{-Qt/2}}{(1-e^{-bt})(1-e^{-t/b})}-
\frac{(Q-2x)^2}{8e^t}-\frac{Q-2x}{t}\biggl).
\ea
Important properties of $\Gamma_b(x)$ are listed as follows
\ba
{}\text{} \text{} \text{functional relation} \quad &
\Gamma_b(x+b)=\sqrt{2\pi}b^{bx-\frac{1}{2}}\Gamma^{-1}(bx)\Gamma_b(x). \label{Ga--feq}\\
{}\text{} \text{} \text{analyticity}\quad &
\Gamma_b(x)\;\text{} \text{} \text{is meromorphic function and it has poles only}\nonumber\\
{}& \text{at}\;\, x=-nb-mb^{-1}, n,m\in\mathbb{Z}^{\geq 0}.\text{} \text{} \text{}
\ea
The further properties is listed in \cite{Sp}.

In terms of functional relation eq.(\ref{Ga--feq}), one can find the residues near by various pole as following
\begin{align}
\label{residuesGamma}
\Gamma_b(x)&={\Gamma_b(Q)\over 2\pi x}+O(x), \text{ }\text{ }\, x\rightarrow 0
\end{align}

\subsection{Double Sine-function}\label{doublesine}
In terms of $\Gamma_b(x)$, the double Sine-function is given as follows
\ba\label{Sb}
S_b(x) = \frac{\Gamma_b(x)}{\Gamma_b(Q-x)}.
\ea
We will use the properties
\begin{align}
\label{Sb1}
\text{Self duality} \quad&
 S_b(x) = S_{b^{-1}}(x) \,, \\
\label{Sb2}
\text{Functional relation} \quad&
 S_b(x + b^{\pm 1}) = 2 \, \sin (\pi b^{\pm 1} x) \, S_b(x) \,,\\
\label{Sb3}
\text{Reflection relation} \quad&
  S_b(x) \, S_b(Q-x) = 1 \,.
\end{align}

The asymptotics behavior of $S_b(x)$ is,
\begin{align}
\quad& S_b(x=x_0+\textup{i}x_1) \sim
\left\{
\begin{aligned}
& e^{\pi  x_1 \left(\frac{b}{2}+\frac{1}{2 b}-x_0\right)}e^{i \pi  \left(\frac{b^2}{12}+\frac{1}{12 b^2}-\frac{b x_0}{2}-\frac{x_0}{2 b}+\frac{x_0^2}{2}-\frac{x_1^2}{2}+\frac{1}{4}\right) }\;\;{\rm for}\;\,
|x|\to \infty,\;\, x_1<0 \,, \\
 &   e^{-\pi  x_1 \left(\frac{b}{2}+\frac{1}{2 b}-x_0\right)}e^{-i \pi  \left(\frac{b^2}{12}+\frac{1}{12 b^2}-\frac{b x_0}{2}-\frac{x_0}{2 b}+\frac{x_0^2}{2}-\frac{x_1^2}{2}+\frac{1}{4}\right) }\;\;{\rm for}\;\,
|x|\to\infty,\;\, x_1>0 \,.
\end{aligned}\right.\label{asymptotic}
\end{align}

We define following functions,
\begin{eqnarray}
\label{otherspecialb}
\Upsilon_b(x) & = & {}{}\frac{1}{\Gamma_b(x)\Gamma_b(Q-x)},{}{}
S_b(x) = {}{}\frac{\Gamma_b(x)}{\Gamma_b(Q-x)},
G_b(x) = {}{}{\rm e}^{-\frac{i\pi}{2}x(Q-x)} S_b(x),
\end{eqnarray}
and, in SFLT, we follow notations from \cite{Fukuda:2002bv} to denote:
\ba\label{SLFTspecialfunction}
\Gamma_{\rm NS}(x) \text{}& = &{}{}
\Gamma_b\left(\frac{x}{2}\right)\Gamma_b\left(\frac{x+Q}{2}\right),
\hskip 1.1cm
\Gamma_{\rm R}(x){}{}
=
\Gamma_b{}{}\left(\frac{x+b}{2}\right)\Gamma_b\left(\frac{x+b^{-1}}{2}\right),
\\[-5pt]
\label{susyspecialdefs}
\\[-5pt]
\nonumber
\Upsilon_{\rm NS}(x){}{}
& = &
\Upsilon_b\left(\frac{x}{2}\right)\Upsilon_b\left(\frac{x+Q}{2}\right),
\hskip 1cm
\Upsilon_{\rm R}(x)
= \Upsilon_b\left(\frac{x+b}{2}\right)\Upsilon_b\left(\frac{x+b^{-1}}{2}\right),
\ea
etc. Using relations, basic properties of these functions can be established easily.
\ba\label{grr}{}
{\Gamma_{\rm NS}(2\alpha){}{}\over \Gamma_{\rm NS}(2\alpha-Q)}&=&W_{\rm NS}(\alpha)\lambda^{{Q-2\alpha\over 2b}}\, ,\\
{\Gamma_{\rm R}(2\alpha){}{}\over \Gamma_{\rm R}(2\alpha-Q)}&=&W_{\rm R}(\alpha)\lambda^{{Q-2\alpha\over 2b}}\, ,
\ea
where $W_{NS}(\alpha)$, $W_{\rm R}(\alpha)$ are defined in (\ref{wn1}) and (\ref{wr1}), and $\lambda=\pi\mu\gamma\left({bQ\over 2}\right)b^{1-b^2}$.
The functions $W_i$ are defined as
\ba\label{wn1}
W_\zt{NS}{}{}{}{}{}{}(\alpha)={2(\pi\mu\gamma(bQ/2))^{-{Q-2\alpha\over 2b}}\pi(\alpha-Q/2)\over \Gamma(1+b(\alpha-Q/2)) \Gamma(1+{1\over b}(\alpha-Q/2))}\, ,
\ea
\ba\label{wr1}
{}{}{}{}W_R(\alpha)={2\pi(\pi\mu\gamma(bQ/2))^{-{Q-2\alpha\over 2b}}\over \Gamma(1/2+b(\alpha-Q/2)) \Gamma(1/2+{1\over b}(\alpha-Q/2))}\,
\ea
In the literature, one can define following equations for convenience,
\ba
{}{}S_{1}(x)\equiv S_{\rm NS}(x)&=&{\Gamma_{\rm NS}(x) \over \Gamma_{\rm NS}(Q-x)}\,{}{} ,\\
{}{}S_{0}(x)\equiv S_{\rm R}(x)&=&{\Gamma_{\rm R}(x) \over \Gamma_{\rm R}(Q-x)}\,{}{} .
\ea
They have the following relations with $S_{NS,R}$ functions:
\ba
{}{}{}{}{S_{\rm NS}(2x)\over S_{\rm NS}(2x-Q)}&=&W_{NS}(x)W_{NS}(Q-x)\,{}{}{}{}{}{} ,\\
{S_{\rm R}(2x){}{}{}{}\over S_{\rm R}(2x-Q)}&=&W_R(x)W_R(Q-x)\,{}{}{}{} .
\ea

In the paper we used following
\begin{itemize}
\item Reflection properties:
\ba
S_{\rm NS}(x)S_{\rm NS}(Q-x) & = &
S_{\rm R}(x)S_{\rm R}(Q-x) = 1 \text{ }\text{ }
\ea
\item Locations of zeroes and poles can be obtained from eq.(\ref{SLFTspecialfunction}):
\ba
S_{\rm NS}(x) & = & 0
\hskip 5mm
\Leftrightarrow
\hskip 5mm
x = Q+ mb+nb^{-1},\text{ }\text{ }
\hskip 5mm
m,n \in {\mathbb Z}_{\geq 0}, \;
m+n \in 2{\mathbb Z},
\\
S_{\rm R}(x) & = & 0
\hskip 5mm \text{ }\text{ }
\Leftrightarrow
\hskip 5mm
x = Q+ mb+nb^{-1},\text{ }\text{ }
\hskip 5mm
m,n \in {\mathbb Z}_{\geq 0}, \;
m+n \in 2{\mathbb Z} +1,\text{ }\text{ }
\\
S_{\rm NS}(x)^{-1} & = & 0
\hskip 5mm
\Leftrightarrow
\hskip 5mm
x = -mb-nb^{-1},
\hskip 5mm
m,n \in {\mathbb Z}_{\geq 0}, \;
m+n \in 2{\mathbb Z},
\\
S_{\rm R}(x)^{-1} & = & 0
\hskip 5mm
\Leftrightarrow
\hskip 5mm
x = -mb-nb^{-1},
\hskip 5mm \text{ }\text{ }
m,n \in {\mathbb Z}_{\geq 0}, \;
m+n \in 2{\mathbb Z} +1.
\ea
\item Basic residue:
\ba
\label{residues}
\lim_{x\to 0}\ x\,S_{\rm NS}(x) & = & \frac{1}{\pi}.
\ea
\end{itemize}

\subsection{{Poles Structure and discrete terms}}\label{discrete}
{ With following the appendix about the LFT in \cite{Cao:2016hvd}, the $4$-point functions of primary operators is:
 \begin{equation}
 \label{intcont11}
 \langle {V}_{\alpha_1}(0){V}_{\alpha_2}(z){V}_{\alpha_3}(1){V}_{\alpha_4}(\infty)\rangle = \int_{i \mathbb{R}+ \frac{Q}{2}} C(\alpha_1,\alpha_2,\alpha_s)C(\alpha_3,\alpha_4, \bar\alpha_s)|{F}_{s}({\Delta_{\alpha_{i=1,2,3,4}}},\Delta_{\alpha_s}, z)|^2d \alpha_s  \,,
\end{equation}}
where $\text{Re}(\alpha_1), \dots, \text{Re}(\alpha_4) = Q / 2$. For $\alpha_i \in (0, Q / 2)$ cases, The eq. (\ref{intcont11}) needs to be extended. The proper way to integration should preserve crossing symmetry with assumptions given by \cite{Zamolodchikov:2006cx}\cite{Belavin:2006ex}\cite{Aleshkin:2016snp}.

The integrand of (\ref{intcont11}), as a function of $\alpha \in \mathbb{C}$, has many poles.
%The conformal block $\mathcal{F}_{\Delta_a}(\{a_i\},z)$, as a function of $\alpha_s$, has poles at
%$ - \frac12L$, where $L \equiv  \left\{b m +  b^{-1} n : m,n= 0,1,2 \dots \right\}$, corresponding to degenerate fields flowing in the internal channel.
The 2 structure constants $C(\alpha_1, \alpha_2, \alpha_s)$ and $C( \alpha_3, \alpha_4,\bar\alpha_s)$ in (\ref{intcont11}) have poles and these poles comes from the zeros of the $\Upsilon$'s in the denominator of the DOZZ formula in LFT\footnote{The similar structure constant in SLFT is given in above appendix \ref{SLFT} and we just list the relevant results.}. These poles are as follows,
\begin{eqnarray}
&\begin{tabular}{|c|c|c|c|c|c|c|c|}
\hline
1 & 2 & 3 & 4 & 5 & 6 & 7 & 8 \\ \hline
 $Q - \alpha_s - L$ &   $\alpha_s  +  L$  &  $2Q - \alpha_s - L$ & $-Q + \alpha_s +  L$ &
 $\alpha_d - L$ & $-\alpha_d + Q + L$  & $-\alpha_d - L$ & $\alpha_d + Q + L$ \\ \hline
\end{tabular} &\nonumber\\
 &\text{where }\alpha_s \equiv \alpha_1 + \alpha_2 \,,\, \alpha_d \equiv \alpha_1 - \alpha_2 \,,\, \text{for LFT}: \text{ }  \text{ }L = \left\{b m +  b^{-1} n : m,n\in \mathbb{Z}^{\geq 0}\right\},&\nonumber\\&  \text{for SLFT NS sector}: \text{ } \text{ } L = \left\{b m +  b^{-1} n : m+n\in 2\mathbb{Z}^{\geq 0}\right\} \,,\, &\nonumber\\&\text{for SLFT R sector}: \text{ } \text{ } L = \left\{b m +  b^{-1} n : m+n\in 2\mathbb{Z}^{\geq 0}+1\right\}&\nonumber\\&&
\label{eqasumdiff11}
\end{eqnarray}
Note that the row $1$ and $2$ are related by $\alpha \mapsto Q - \alpha$ symmetry and so do $3$ and $4$, \textit{etc}. The poles coming from $C( \alpha_2, \alpha_3, \bar\alpha)$ are got by replacing $1,2 \to 3,4$ in the above equations.

If $\text{Re}(\alpha_i) = \frac{Q}{2}$, the real part of the poles belongs to the intervals $(-\infty, 0] \cup [Q, +\infty)$ and the intervals do not intersect with the integration contour $Q/2 + i R$ (\ref{intcont11}). When $\text{Re}(\alpha_s) = \text{Re}(\alpha_1+\alpha_2)$ starts to decrease from $Q$ into the interval $(0,Q)$, the poles start to move on the plane $\alpha_s$. One can show that only the rows $1$ and $2$ may cross the line $Q/2 + i R$. When $\text{Re}(\alpha_s)$ decreases to $Q / 2$, row $1$ crosses the line from its left, and row $2$ cross the line from it right. As $\text{Re}(\alpha_s)$ further decreases, several poles from those rows in the table will have crossed the line. These poles are:
\begin{align}
 P_+ \equiv \left\{ x \in Q - (\alpha_1 + \alpha_2) - L : \text{Re}(x) \in  (Q/2, Q) \right\} \,,\, P_- \equiv \left\{ x \in \alpha_1 + \alpha_2 + L : \text{Re}(x) \in  (0, Q / 2) \right\}
 \,. \label{eqincludepoles11}
\end{align}
To extend analytically the integral (\ref{intcont11}), the integration contour has to be deformed to avoid the poles from crossing it. Using Cauchy formula, this amounts to adding $ \pm 2\pi i$ times the residues of the integrand of (\ref{intcont11}) at points in $P_\pm$, respectively. These terms are the so called discrete terms \cite{Belavin:2006ex} \cite{Zamolodchikov:2006cx}\cite{Aleshkin:2016snp} known in the literature.

Using the $\alpha_s \mapsto Q - \alpha_s$ symmetry of $\Upsilon$, the contribution of $P_+$ equals that of $P_-$. Then, the poles from $\alpha_{2,3}$ can be similarly treated. For LFT $4$-point function for values of $\text{Re}(\alpha_i)\in (0,Q/2) $, the resulting form is as follows \cite{Cao:2016hvd}:
  \begin{eqnarray}
  \label{intpdis11}
  \langle {V}_{\alpha_1}(0){V}_{\alpha_2}(z){V}_{\alpha_3}(1){V}_{\alpha_4}(\infty)\rangle = \int_{\frac{Q}{2} + i \mathbb{R}} C(\alpha_1,\alpha_2,\alpha_s)C(\alpha_3,\alpha_4,\bar\alpha_s)|{F}_{s}(\Delta_{\alpha_i},\Delta_{\alpha_s}, z)|^2 d \alpha_s +\nonumber \\
  -   2\sum_{p \in P_-}  \;\text{Res}_{\alpha_s \to p} \left[C(\alpha_1,\alpha_2,\alpha_s) C(\alpha_3,\alpha_4,\bar\alpha_s)\right]  |{{F}_{s}(\Delta_{\alpha_i}, \Delta_{\alpha_s}, z)}|^2 + [(1,2) \leftrightarrow (3,4)] \,.
  \end{eqnarray}

In this paper, we will use asymptotic form of conformal block as $z\to 0, \text{or} \text{ } z \to 1$. The $z\to 0$ series expansion of conformal block is:
\begin{equation}
\label{cbexp11}
{F}_{s}(\Delta_{\alpha_i},\Delta_{\alpha_s},z) = z^{-\Delta_{\alpha_1}-\Delta_{\alpha_2}+\Delta_{\alpha_s}} \left(1+\frac{(\Delta_{\alpha_2}-\Delta_{\alpha_1}+ \Delta_{\alpha_s})(\Delta_{\alpha_3}-\Delta_{\alpha_4}+ \Delta_{\alpha_s})}{2\Delta_{\alpha_s}} z + O(z^2)\right) \,.
\end{equation}

Once $s$-channel blocks are known, $t$-channel blocks can be obtained by a permutation of the arguments, with taking global conformal symmetry into account and this becomes \cite{Ribault:2014hia}
\begin{align}
{F}_{t}(\Delta_{\alpha_i},\Delta_{\alpha_t}, z) = (1-z)^{\Delta_2+\Delta_3-\Delta_1-\Delta_4}{F}_{s}(\Delta_{\alpha_i},\Delta_{\alpha_s},1-z)\ .
\end{align}
In this paper, to compute the dominant asymptotic behavior of (\ref{eqincludepoles11}) as $z\to 0$ or $z\to 1$ is very important. We need to consider the internal charges $\alpha_s \in P_+ \cup (Q/2 + i \mathbb{R})$ involved, and find the {smallest} scaling dimension $\Delta_\alpha = \alpha(Q - \alpha)$. In this paper, we will make use of these details to obtain REE. Some further calculations have been given in subsection \ref{appendix2}.

\subsection{To calculate the dominant contribution in early time}\label{appendix2}
In this section, we would like to show the some details about how to do the early time integral appeared in four point function in LFT.
The early time limit $(z,\bar z)\to (0,0)$ is a short distance limit, or equivalently $z_1 \to z_2$ or $z_3 \to z_4$.
Hence one can insert a corresponding OPE \cite{Belavin:2006ex}\cite{Aleshkin:2016snp} in eq.(\ref{intcont11}). If the OPE would be given by a sum
as in rational CFT¡¯s, the dominant contribution trivially would be realized by the
contribution of the r.h.s. operator with lowest conformal dimension. In LFT, we have an integral as given in formula (1.10) of \cite{Belavin:2006ex}.

For $\text{Re}(\alpha_1+\alpha_2)>Q/2$ \cite{Belavin:2006ex}, the integral is
\bea
\langle {V}_{\alpha_1}(0){V}_{\alpha_2}(z){V}_{\alpha_1}(1){V}_{\alpha_2}(\infty)\rangle \underset{(z,\bar z)\to (0,0)}{\simeq} \int_0^{\infty} (z\bar z)^{P^2} f(P) dP
\eea
whose asymptotics for $z\to 0$ is
\bea\label{seriese}
\sqrt{\pi\over -\log z}{f(0)\over 2}+{1\over 2\log z}f'(0)+{\sqrt{{\pi}}\over 8 (-\log z)^{3\over2}}f''(0)+...
\eea
One can apply to following OPE \cite{Belavin:2006ex}
\bea
V_{\alpha_1}(0)V_{\alpha_2}(z)=\int'{d P\over 4 \pi}C(\alpha_1,\alpha_2,\alpha_s) (z\bar z)^{\Delta_{Q/2+ i P}-\Delta_{\alpha_1}-\Delta_{\alpha_2}}[V_{Q/2+i P}(0)]
\eea
The integration contour here is the real axis if $\alpha_1$ and $\alpha_2$ are in the ¡°basic domain¡±$|Q/2-\text{Re}(\alpha_1)|+|Q/2-\text{Re}(\alpha_2)|<Q/2$.
In this case, one can find that
\bea
{1\over 4\pi}\sqrt{\pi\over -\log(z\bar z)}(z\bar z)^{{Q^2\over 4}-\Delta_1-\Delta_2}C(\alpha_1,\alpha_2,{Q\over 2})V_{Q/2}(0)+...
\eea

But we have to take into account that $C(\alpha_1,\alpha_2,{Q\over 2})=0$. The function $f(P)$ is $C(\alpha_1,\alpha_2,{Q\over 2}+iP)C(\alpha_1,\alpha_2,{Q\over 2}- iP)$. Due to $f¡ä(0) = 0$ and the first nonvanishing term is the one with $f''(0)$ which is then
\bea
\langle {V}_{\alpha_1}(0){V}_{\alpha_2}(z){V}_{\alpha_1}(1){V}_{\alpha_2}(\infty)\rangle \underset{(z,\bar z)\to (0,0)}{\simeq} {\sqrt{{\pi}}\over 8 }f''(0) (z\bar z)^{{Q^2\over 4}-\Delta_1-\Delta_2}(-\log(z\bar z))^{-{3\over 2}}
\eea
This conclusion corresponds to a statement by Seiberg in \cite{Seiberg:1990eb}, see comment after his
eq. (4.15).\footnote{The fact that author writes $Q^2/8$ instead of $Q^2/4$ due to different
normalisations.}

For $\alpha_1$ and $\alpha_2$ stay outside of ¡°basic domain¡±\cite{Belavin:2006ex}, we can use following OPE
\bea\label{discretetermLFT}
V_{\alpha_1}(0)V_{\alpha_2}(z)&=& {1\over 2} (z\bar z)^{-2\alpha_1\alpha_2}[V_{\alpha_1+\alpha_2}(0)]+ {1\over 2} (z\bar z)^{-2\alpha_1\alpha_2}S(\alpha_1+\alpha_2)[V_{Q-\alpha_1-\alpha_2}(0)]\nonumber\\&+&{1\over 2}\int{d P\over 4 \pi}C(\alpha_1,\alpha_2,\alpha_s) (z\bar z)^{\Delta_{Q/2+ i P}-\Delta_{\alpha_1}-\Delta_{\alpha_2}}[V_{Q/2+i P}(0)]
\eea

The integral has again an asymptotics as above eq.(\ref{seriese}). The first two terms in eq.(\ref{discretetermLFT}) are so called discrete terms. But now the more dominant term is given by the contribution of the discrete term, i.e.
\bea
&&\langle {V}_{\alpha_1}(0){V}_{\alpha_2}(z){V}_{\alpha_1}(1){V}_{\alpha_2}(\infty)\rangle\nonumber\\&\underset{(z,\bar z)\to (0,0)}{\simeq}& {C(\alpha_1,\alpha_2, \alpha_1+\alpha_2)C(\alpha_1,\alpha_2,Q-\alpha_1-\alpha_2)} (z\bar z)^{\Delta_{\alpha_1+\Delta_{\alpha_2}}-\Delta_1-\Delta_2}V_{\alpha_1+{\alpha_2}}(0)\nonumber\\
\eea
Depending on the value of $\alpha_1$ and $\alpha_2$ other discrete terms can take over,
as discussed in \cite{Belavin:2006ex}.

To calculate the late time limit of $\langle {V}_{\alpha_1}(0){V}_{\alpha_2}(z){V}_{\alpha_1}(1){V}_{\alpha_2}(\infty)\rangle$, we have to use conformal boostrap equations firstly and following the above procedures in this subsection. We have present the main results and will not repeat all the details here.

\subsection{The Fusion matrix in Liouville field theory}\label{fusionLFT}
In this subsection, we will see how to associate with the quantum dimension defined in LFT.
For late time limit of the second REE, the fusion matrix element $F^L_{{{Q/2}},{Q/2}}\big[ {}_{\xa}^{\bar\xa} {}_{\bar{\xa}}^{ \xa} \big]$ will be presented. This matrix element can not be identify the quantum dimension in LFT. With following the definition of quantum dimension \cite{McGough:2013gka} LFT, we will show the $F^L_{{{0}},{0}}\big[ {}_{\xa}^{\bar\xa} {}_{\bar{\xa}}^{ \xa} \big]$ will be the quantum dimension.

Just to follow the convention in \cite{Vartanov:2013ima}\cite{Teschner:2001rv}\cite{Ponsot:1999uf}, we introduce ${F_{\xa_s \xa_t}^{\rm L} \big[ {}_{\xa_4}^{\xa_3} {}_{\xa_1}^{\xa_2} \big]}$ as followings
\begin{align} \label{norm}
{F_{\xa_s \xa_t}^{\rm L} \big[ {}_{\xa_4}^{\xa_3} {}_{\xa_1}^{\xa_2} \big]}
&={\frac{N(\xa_s,\alpha_2,\alpha_1)N(\alpha_4,\alpha_3,\xa_s)}{N(\xa_t,\alpha_3,\alpha_2) N(\alpha_4,\xa_t,\alpha_1)} }
{F_{\xa_s \xa_t}^{\rm PT} \big[ {}_{\xa_4}^{\xa_3} {}_{\xa_1}^{\xa_2} \big]}\,,
\end{align}
where
\begin{align}\label{N-def}
&N(\xa_3,\xa_2,\xa_1)\,\,\nonumber\\
=&\,\frac{\xG_b(2Q-2\xa_3)\xG_b(2\xa_2)\xG_b(2\xa_1)\xG_b(Q)}
{\xG_b(2Q-\xa_1-\xa_2-\xa_3)\xG_b(Q-\xa_1-\xa_2+\xa_3)
\xG_b(\xa_1+\xa_3-\xa_2)\xG_b(\xa_2+\xa_3-\xa_1)}\,.
\end{align}
The $b$-6$j$ symbol has the following explicit form
\begin{align}
\label{6j1}
{F_{\xa_s \xa_t}^{\rm PT} \big[ {}_{\bar \xa_4}^{\xa_3} {}_{\xa_1}^{\xa_2} \big]} = &
\;\frac{S_b(\alpha_2+\alpha_s-\alpha_1) S_b(\alpha_t+\alpha_1-\alpha_4)}{S_b(\alpha_2+\alpha_t-\alpha_3) S_b(\alpha_s+\alpha_3-\alpha_4)}\,|S_b(2\xa_t)|^2 \nonumber\\
& \times  \int_{\CC}du\;
S_b(-\alpha_2 \pm (\alpha_1-Q/2)+u)
S_b(-\alpha_4 \pm (\alpha_3-Q/2) +u) \nonumber \\[-.5ex] &
\hspace{1.125cm}\times
S_b(\alpha_2 + \alpha_4 \pm (\alpha_t-Q/2) - u)
S_b(Q \pm (\alpha_s - Q/2) - u) \,,
\end{align}
where the following notation has been used
$
S_b(\alpha \pm u) :=
S_b(\alpha+u)
S_b(\alpha-u).
$ The function $S_b(x)$ is defined by eq.(\ref{Sb}) in appendix \ref{doublesine}.
The integral can be performed using the identity
\begin{eqnarray}\label{Corollary 1}
\int_{i\BR} % -\textup{i} \infty}^{\textup{i} \infty}
dz\; \prod_{i=1}^3 S_b(\mu_i-z)
S_b(\nu_i+z) = \prod_{i,j=1}^3 S_b(\mu_i+\nu_j),
\end{eqnarray}
where the balancing condition is $\sum_{i=1}^3 \mu_i+\nu_i = Q$.
We are interested in the $F_{\xa_s \xa_t}^{\rm PT} \big[ {}_{\xa}^{\bar{\xa}} {}_{\bar{\xa}}^{\xa} \big]$ which has been shown in section \ref{2REELFT}. In terms of notation of \cite{Teschner:2012em}, the $6j$ symbols correspond to $F_{\xa_s \xa_t}^{\rm PT} \big[ {}_{\alpha_4}^{\alpha_3} {}_{\alpha_1}^{\alpha_2} \big] =\big\{\,{}^{\alpha_1}_{\alpha_3}\,{}^{\alpha_2}_{{\alpha}_4}\,|\,{}^{\alpha_s}_{\alpha_t}\big\}_b^{\rm an}$ and canonical $6j$ symbols are defined as
\begin{equation} \label{normlized}
\big\{\,{}^{\alpha_1}_{\alpha_3}\,{}^{\alpha_2}_{{\alpha}_4}\,|\,{}^{\alpha_s}_{\alpha_t}\big\}_b^{}={
\frac{M(\alpha_s,\alpha_2,\alpha_1)M(\alpha_4,\alpha_3,\alpha_s)}{M(\alpha_t,\alpha_3,\alpha_2) M(\alpha_4,\alpha_t,\alpha_1)} }
\big\{\,{}^{\alpha_1}_{\alpha_3}\,{}^{\alpha_2}_{{\alpha}_4}\,|\,{}^{\alpha_s}_{\alpha_t}\big\}_b^{\rm an}\,.
\end{equation}
with
\ba
 M(\alpha_3,\alpha_2,\alpha_1)  &=
\Big(S_b(2Q-\alpha_1-\alpha_2-\alpha_3) S_b(Q-\alpha_1-\alpha_2+\alpha_3)
\nonumber\\ &S_b(\alpha_1+\alpha_3-\alpha_2) S_b(\alpha_2+\alpha_3-\alpha_1)\Big)^{-\frac{1}{2}}\label{normalizationfactorF}.
\ea
With following relation given in \cite{Teschner:2012em},
\ba\label{6j-moddeff}
\big\{\,{}^{\alpha_1}_{\alpha_3}\;{}^{\alpha_2}_{\alpha_4}\;{}^{\alpha_s}_{\alpha_t}\big\}_b
=
\big\{\,{}^{\alpha_1}_{\alpha_3}\,{}^{\alpha_2}_{\bar\alpha_4}\,|\,{}^{\alpha_s}_{\alpha_t}\big\}_b\,,
\qquad
\bar\alpha_4:=Q-\alpha_4.
\ea
one can obtain that
\ba\label{PrefactorforFusion}
F_{\xa_s \xa_t}^{\rm PT} \big[ {}_{\alpha_4}^{\alpha_3} {}_{\alpha_1}^{\alpha_2} \big]={ M(\alpha_4,\alpha_t,\alpha_1) M(\bar{\alpha_4},\alpha_3,\alpha_s)\over  M(\bar{\alpha_4},\alpha_t,\alpha_1) M(\alpha_4,\alpha_3,\alpha_s)}F_{\xa_s \xa_t}^{\rm PT} \big[ {}_{\bar{\alpha_4}}^{\alpha_3} {}_{\alpha_1}^{\alpha_2} \big].
\ea
Then
\ba {F_{\xa_s \xa_t}^{\rm L} \big[ {}_{\xa}^{\bar\xa} {}_{\bar\xa}^{\xa} \big]}&=&{\xG_b(2\alpha_s)\over \xG_b(\alpha_s)}{\xG_b(\alpha_t)\over \xG_b(2\alpha_t)}\frac{\Gamma _b\left(2 Q-2 \alpha _s\right) \Gamma _b\left(Q-\alpha _t\right){}^3}{\Gamma _b\left(2 \alpha -\alpha _s\right) \Gamma _b\left(Q-\alpha _s\right){}^3}\nonumber\\
&&\frac{\Gamma _b\left(2 \alpha -\alpha _t\right) \Gamma _b\left(-2 \alpha +Q+\alpha _t\right) \Gamma _b\left(2 \alpha -Q+\alpha _t\right){}^2}{\Gamma _b\left(-2 \alpha +Q+\alpha _s\right) \Gamma _b\left(2 \alpha -Q+\alpha _s\right){}^2 \Gamma _b\left(2 Q-2 \alpha _t\right)}\nonumber\label{Normalizationfactor1}\\
&&{\xG_b(\alpha_t)\over \xG_b(\alpha_s)}\frac{\Gamma _b\left(Q-\alpha _s\right) \Gamma _b\left(2 \alpha -Q+\alpha _s\right) \Gamma _b\left(-2 \alpha +2 Q-\alpha _t\right)}{\Gamma _b\left(-2 \alpha +2 Q-\alpha _s\right) \Gamma _b\left(Q-\alpha _t\right) \Gamma _b\left(2 \alpha -Q+\alpha _t\right)}\nonumber\label{Normalizationfactor2}\\
&&\frac{\Gamma _b\left(\alpha _s\right) \Gamma _b\left(-2 \alpha +Q+\alpha _s\right) \Gamma _b\left(2 \alpha -Q+\alpha _s\right)}{\Gamma _b\left(2 \alpha -\alpha _s\right) \Gamma _b\left(Q-\alpha _s\right) \Gamma _b\left(-2 \alpha +2 Q-\alpha _s\right)}|S_b(2\xa_t)|^2\label{Normalizationfactor3}
\ea
The factor in the second line of (\ref{Normalizationfactor3}) comes from 4 normalization factors $N(\alpha_3,\alpha_2,\alpha_1)$ in eq.(\ref{norm}), the factor in the third line of (\ref{Normalizationfactor3}) is from 4 factors associated with $M(\alpha_3,\alpha_2,\alpha_1)$ in eq.(\ref{PrefactorforFusion}) and the factor in the last line of (\ref{Normalizationfactor3}) is mainly from $F_{\xa_s \xa_t}^{\rm PT} \big[ {}_{\alpha_4}^{\alpha_3} {}_{\alpha_1}^{\alpha_2} \big]$ eq.(\ref{6j1}). We have already made use of eq.(\ref{Corollary 1}) to do the $u$ integration to obtain the simple expression (\ref{Normalizationfactor3}). From (\ref{Normalizationfactor3}) and meromorphic property of $\Gamma _b(\alpha)$ shown in appendix, one can see that there is no pole structure in ${F_{\xa_s \xa_t}^{\rm L} \big[ {}_{\xa}^{\bar\xa} {}_{\bar\xa}^{\xa} \big]}$ for $\xa_s \to 0$.
We then obtain the following expansion near $\xa_s \to 0$,
\ba\label{FPTspecial}
{F_{0\xa_t}^{\rm L} \big[ {}_{\xa_2}^{\xa_2} {}_{\xa_1}^{\xa_1} \big]}&=&{1\over 2}|S_b(2\xa_t)|^2
\frac{ \Gamma _b\left(\alpha _t\right){}^2}{\Gamma _b\left(2\alpha _t\right){}}\frac{\Gamma _b(2 Q) \Gamma _b\left(Q-\alpha _t\right){}^2 \Gamma _b\left(-2 \alpha +2 Q-\alpha _t\right)}{\Gamma _b(Q){}^3 \Gamma _b(2 Q-2 \alpha ){}^2}\nonumber\\&&\frac{\Gamma _b\left(2 \alpha -\alpha _t\right) \Gamma _b\left(-2 \alpha +Q+\alpha _t\right) \Gamma _b\left(2 \alpha -Q+\alpha _t\right)}{\Gamma _b(2 \alpha ){}^2 \Gamma _b\left(2 Q-2 \alpha _t\right)}
\ea

The factor $|S_b(2\xa_t)|^2$ can be taken care of using
\[|S_b(\alpha)|^2=-4\sin\pi b(2\alpha-Q)\sin\pi \iv b(2\alpha-Q),\]
but we will temporarily keep it. The only divergence is from the simple pole of $\xG_b(\xa_t)$ eq.(\ref{residuesGamma}),
\be
\Gamma_b(x)\sim {\Gamma_b(Q)\over 2\pi x}\,,
\ee
and the residue is given by simple pole of ${F_{0\xa_t\to 0}^{L} \big[ {}_{\xa}^{\bar\xa} {}_{\bar\xa}^{\xa} \big]}$
%\textcolor{blue}
{\be
{F_{0\xa_t\to 0}^{L} \big[ {}_{\xa}^{\bar\xa} {}_{\bar\xa}^{\xa} \big]} = \frac{1}{2 \pi }{|S_b(\alpha)|^2 }.\frac{\Gamma _b(Q-2 \alpha ) \Gamma _b(2 \alpha -Q)}{\Gamma _b(2 \alpha ) \Gamma _b(2 Q-2 \alpha )}=\frac{1}{2 \pi }{|S_b(\alpha)|^2 }.\frac{ S _b(2 \alpha -Q)}{S_b(2 \alpha ) }
\ee}
With the help of the following identity eq.(\ref{Sb2})
\be
S_b(x + b^{\pm 1}) = 2 \, \sin (\pi b^{\pm 1} x) \, S_b(x)\,,
\ee
we can express the residue as (where we also use $S_b(x) = S_{\iv b}(x)$)
\ba\label{quantumfourbar}
{F_{0\xa_t \to 0} \big[ {}_{\xa}^{\bar\xa} {}_{\bar\xa}^{\xa} \big]} & = & \frac {1} {2\pi \xa_t} \frac{|S_b(2\xa_t)|^2 S_b(2 \alpha -Q)}{4 \sin [\pi b (2\xa-Q+\frac 1 b)] \sin [\pi \iv b (2\xa-Q)] S_b(2 \alpha -Q)} \nn
& = & \frac {1} {2\pi \xa_t} \frac{\sin\pi b Q \sin\pi \iv b Q}{\sin [\pi b (2\xa-Q)] \sin [\pi \iv b (2\xa-Q)]} \,.
\ea

With comparing the definition of quantum dimension \cite{McGough:2013gka} in LFT, we will show the $F^L_{{{0}},{0}}\big[ {}_{\xa}^{\bar\xa} {}_{\bar{\xa}}^{ \xa} \big]$ will be the quantum dimension.

\subsection{The Fusion matrix in super Liouville field theory}\label{fusionSLFT}
Similar to the subsection \ref{fusionLFT}, we would like to comment on quantum dimension and fusion matrix presented in the second REE in SLFT.
Generically, the fusion matrices take the following form $i,j=1,2$ which correspond to parity of $e$ and $o$ respectively \cite{Poghosyan:2016kvd}
\ba
&&F_{\alpha_s,\alpha_t}\left[\begin{array}{cc}
\alpha_3& \alpha_2\\
\alpha_4& \alpha_1 \end{array}\right]^i_j\label{FusionmatrixinSLFT}\\ \nonumber
&=&{\Gamma_i(2Q-\alpha_t-\alpha_2-\alpha_3)\Gamma_i(Q-\alpha_t+\alpha_3-\alpha_2)\Gamma_i(Q+\alpha_t-\alpha_2-\alpha_3)
\Gamma_i(\alpha_3+\alpha_t-\alpha_2)\over
\Gamma_j(2Q-\alpha_1-\alpha_s-\alpha_2)\Gamma_{j}(Q-\alpha_s-\alpha_2+\alpha_1)\Gamma_j(Q-\alpha_1-\alpha_2+\alpha_s)
\Gamma_{j}(\alpha_s+\alpha_1-\alpha_2)}\\  \nonumber
&\times&{\Gamma_i(Q-\alpha_t-\alpha_1+\alpha_4)\Gamma_i(\alpha_1+\alpha_4-\alpha_t)\Gamma_i(\alpha_t+\alpha_4-\alpha_1)
\Gamma_i(\alpha_t+\alpha_1+\alpha_4-Q)\over
\Gamma_{j}(Q-\alpha_s-\alpha_3+\alpha_4)\Gamma_j(\alpha_3+\alpha_4-\alpha_s)\Gamma_{j}(\alpha_s+\alpha_4-\alpha_3)
\Gamma_j(\alpha_s+\alpha_3+\alpha_4-Q)}\\  \nonumber
&&\times {\Gamma_{\rm NS}(2Q-2\alpha_s)\Gamma_{\rm NS}(2\alpha_s)\over \Gamma_{\rm NS}(Q-2\alpha_t)\Gamma_{\rm NS}(2\alpha_t-Q)}
{1\over i}\int_{-i\infty}^{i\infty}d\tau J_{\alpha_s,\alpha_t}\left[\begin{array}{cc}
\alpha_3& \alpha_2\\
\alpha_4& \alpha_1 \end{array}\right]^i_j\, ,
\ea
We will consider $i=j=1$, which gives the fusion matrix for NS sector. In this case, we have
\ba
&&J_{\alpha_s,\alpha_t}\left[\begin{array}{cc}
\alpha_3& \alpha_2\\
\alpha_4& \alpha_1 \end{array}\right]^1_1\\ \nonumber
&=& {S_\zt{NS}(Q+\tau-\alpha_1)S_{\rm NS}(\tau+\alpha_4+\alpha_2-\alpha_3)S_\zt{NS}(\tau+\alpha_1)S_\zt{NS}(\tau+\alpha_4+\alpha_2+\alpha_3-Q)\over
S_\zt{NS}(Q+\tau+\alpha_4-\alpha_t)S_\zt{NS}(\tau+\alpha_4+\alpha_t)S_\zt{NS}(Q+\tau+\alpha_2-\alpha_s)S_\zt{NS}(\tau+\alpha_2+\alpha_s)}\\ \nonumber
&+&{S_{R}(Q+\tau-\alpha_1)S_{R}(\tau+\alpha_4+\alpha_2-\alpha_3)S_{R}(\tau+\alpha_1)S_{R}(\tau+\alpha_4+\alpha_2+\alpha_3-Q)\over
S_{R}(Q+\tau+\alpha_4-\alpha_t)S_{R}(\tau+\alpha_4+\alpha_t)S_{R}(Q+\tau+\alpha_2-\alpha_s)S_{R}(\tau+\alpha_2+\alpha_s)}\, ,
\ea
where $S_{NS,R}$ are defined by eq.(\ref{SLFTspecialfunction}).
In terms of explicit form of $F_{\alpha_s,\alpha_t}\left[\begin{array}{cc}
\alpha_3& \alpha_2\\
\alpha_4& \alpha_1 \end{array}\right]$ in eq.(\ref{FusionmatrixinSLFT}), we can show that \ba F_{\alpha_s,\alpha_t}\left[\begin{array}{cc}
\bar{\alpha}& \alpha\\
\alpha& \bar{\alpha} \end{array}\right]_1^{1}=F_{\alpha_s,\alpha_t}\left[\begin{array}{cc}
\alpha& \alpha\\
\alpha& \alpha \end{array}\right]_1^{1}. \ea

Alternatively, it can be shown that for $\xa_s = 0$, the fusion matrix becomes
\be
\label{Fusionmatrixpole}
F_{0,\alpha_t}\left[\begin{array}{cc}
\alpha_3& \alpha_1\\
\alpha_3& \alpha_1 \end{array}\right]^1_1=C_\zt{NS}(\alpha_t,\alpha_1,\alpha_3){W_\zt{NS}(Q)W_\zt{NS}(\alpha_t)
\over \pi W_\zt{NS}(Q-\alpha_1)W_\zt{NS}(Q-\alpha_3)}\, .
\ee Where $W_\zt{NS}$ is defined in (\ref{wn1}) and (\ref{wr1}).
It is not difficult to see that near $\xa_t \sim 0$ (and $\xa_1 = \xa_3 = \xa$), the DOZZ function has
a single pole
\ba\label{dozzpole}
C_\zt{NS}(\alpha_t,\alpha,\alpha)&\sim &\lambda^{(Q-2\alpha)/b}\frac {\Upsilon'_\zt{NS}(0)\Upsilon_\zt{NS}(2\alpha_t)\Upsilon_\zt{NS}(2\alpha)}{
\Upsilon_\zt{NS}(2\alpha-Q)\Upsilon_\zt{NS}(\alpha_t)^2}\, ,\nn
& \sim & \frac {\Upsilon'_\zt{NS}(0)\Upsilon_\zt{NS}(2\alpha_t)W_\zt{NS}(Q-\alpha)}{
W_\zt{NS}(\alpha)\Upsilon^2_\zt{NS}(\alpha_t)} \sim \frac {2W_\zt{NS}(Q-\alpha)}{
W_\zt{NS}(\alpha)\pi \alpha_t}\,,
\ea
where in the second line we use
\be
\Gamma_{\rm NS}(x)\sim {\Gamma_{\rm NS}(Q)\over \pi x} \Rightarrow \Upsilon_\zt{NS}(x)
\sim \frac {\pi x}{\Gamma^2_{\rm NS}(Q)}\, .
\ee

In the second line of \er{dozzpole} we use
\be
{\Upsilon_{\rm NS}(2x)\over \Upsilon_{\rm NS}(2x-Q)}={\cal G}_\zt{NS}(x)\lambda^{-{Q-2x\over b}}={W_\zt{NS}(Q-x)\over W_\zt{NS}(x)}\lambda^{-{Q-2x\over b}}\, .
\ee
We also need the values of the derivative $\Upsilon_{\rm NS}'(0)$
\be
\Upsilon_{\rm NS}'(0)={\pi \over \Gamma_{\rm NS}^2(Q)}\, .
\ee
We have made use of definition of $\Gamma_{\rm NS}, \Gamma_{\rm R}$ given in eq.(\ref{SLFTspecialfunction}).
Substituting \er{dozzpole} back into \er{Fusionmatrixpole} and with the help
of the following identity
\be
W_\zt{NS}(x)W_\zt{NS}(Q-x) = -4\sin\pi b(x-Q/2)\sin\pi {1\over b}(x-Q/2)\, ,
\ee
we obtain the pole structure of the fusion matrix
\be
 F_{0,\alpha_t}\left[\begin{array}{cc}
\alpha & \alpha\\
\alpha & \alpha \end{array}\right]^1_1 = \frac 2 {\pi^2 \xa_t}\frac {\sin\frac \pi 2 bQ\sin\frac \pi 2 \iv b Q}{\sin\pi b(\xa-Q/2)\sin\pi \iv b(\xa-Q/2)}\,.
\ee
So we can again relate the entanglement entropy due to the the local operator to its quantum dimension
\be
\zt{Res}_{\xa_t = 0} F_{00} \sim  \sin\pi b(\xa-Q/2)\sin\pi \iv b(\xa-Q/2)\,.
\ee
Similar to the situation in LFT, fusion matrix element for $F^{SL}_{{{Q/2}},{Q/2}}\big[ {}_{\xa}^{\xa} {}_{\xa}^{\xa} \big]^{e}_{e}$ presented in the second REE (\ref{integralsuper}) in SLFT can not be identify as quantum dimension.

  \end{document}